\documentstyle[prb,aps]{revtex}

\def\lsim{\mathrel{\rlap{\lower4pt\hbox{\hskip1pt$\sim$}}
    \raise1pt\hbox{$<$}}}         
\def\gsim{\mathrel{\rlap{\lower4pt\hbox{\hskip1pt$\sim$}}
    \raise1pt\hbox{$>$}}}         
\def\be{\begin{equation}}
\def\ee{\end{equation}}
\def\bq{\begin{eqnarray}}
\def\eq{\end{eqnarray}}
\def\bqn{\begin{eqnarray*}}
\def\eqn{\end{eqnarray*}}

\begin{document}

\draft
\title{A coarse-grained, ``realistic'' model for protein
folding}
\author{Pierpaolo  Bruscolini }
\address{Dipartimento di Fisica Teorica, Universit\`a di Torino, \\
v. Giuria 1, 10125 Torino, ITALY}

\date{\today}
\maketitle
\begin{abstract}
A phenomenological model hamiltonian to describe 
the folding of a protein with any given sequence is proposed.
The protein is thought of as a collection of pieces of helices; as a
consequence  
its configuration space increases with the number of secondary
structure elements rather than with the number of residues. 
The hamiltonian presents both local (i.e. single helix, accounting for
the stiffness of the chain) and non local (interactions between
hydrophobically--charged helices) terms, and is expected to provide
a first tool for studying the folding of real proteins. 
The partition function for a 
simplified, but by no means trivial,     
version of the model is calculated almost completely in an analytical way.
The latter simplified model is also applied to the study of 
a synthetic protein, and
some preliminary results are shown.
\end{abstract}

\section{Introduction}
Protein folding is one of the most challenging problems in molecular
biology, and many theoretical models have been proposed, aimed to
explain the thermodynamics as well as the kinetics of this process.

It is experimentally known~\cite{anfi73,anfi75,chot92} that a
protein, under proper solvent and temperature conditions,
folds from a denatured random-shaped  state to its ``native''
state, which is characterized only by the amino acid sequence ("the primary
structure"), and
does not depend on the initial state. 

The  experimental data for the folding process support the so-called 
"molten-globule" picture~\cite{chpa94}: 
folding of small single-domained proteins in
proper
solvent conditions would start with a rapid collapse from a coil state
to a compact one, which is not unique, but  is in metastable equilibrium
with several other compact states. 
The protein would appear at this point as a molten globule, not presenting 
a definite shape. The latter  diffuses among the various conformations
until it finds its way across the free-energy barrier (probably unique)
separating it from the most stable (native) state, and eventually would 
reach the native state.
Because of the cooperativity of the process, folding appears from a
thermodynamical point of view as an all-or-none transition, similar to
a first-order  phase transition in infinite systems \cite{crei94}.

The understanding of the physics involved in protein folding has greatly
 profited by ideas and techniques coming from statistical mechanics of
disordered systems 
and
random  energy models (R.E.M.) \cite{derr80}. As far as thermodynamics
is concerned,
the number and the complexity of interactions in which residues are
involved, as well as the fact that functionally similar proteins may
have somewhat different sequences (for instance, lysozymes of different
species), have suggested to approach the folding problem by means of an
analogy with R.E.M. \cite{brwo87}.  Analytic results \cite{shgu190}
predict  a glass-transition when
the probability distribution of the couplings is broad enough (that is,
when the residues behaves very differently from one another). Several
studies have been carried out on short model heteropolymers on a
lattice (both with random and with specified interactions), where  a
preliminary complete enumeration always allows us to find the energy of any
configuration.  These studies have revealed some important requirements
that a sequence should fulfil in order to be a good folder (gap in the
spectrum, non-degeneracy of the ground state \cite{sasha94},
particular values of the ratio between collapse and folding
temperatures \cite{klth96}). Debate is still  open  on the
relative importance of these properties for characterizing good
sequences \cite{thkl96}.
However, what the 
analysis has assessed is that these requirements are not typical
of  random
sequences, so that a true protein cannot be considered a random
heteropolymer, as long as the feature of being able to fold to a
stable and unique native state is concerned.

Lattice models have also been employed to study the kinetic aspects of 
the folding process \cite{shfaa91,soon95,klth96}, under the
hypothesis that  they do not  depend on 
microscopic details of the dynamics. One still gets thus meaningful
results when using fictitious Monte Carlo dynamics instead of the true
one (which, of course, cannot be easily implemented  on a lattice). The
above assumption seems reasonable, since the predictions of
various  diffusive regimes and relaxation times, that come out of
these studies, are in good qualitative agreement with phenomenology
\cite{thir95}.  
Besides, these works have revealed that kinetic accessibility of the
native state from a generic initial condition is as important as the
already mentioned   thermodynamical requirements, for a sequence to be a
good folder. As a consequence, it is commonly believed that real
proteins present a rough but funnel-shaped free-energy landscape,
which provides an overall bias towards the native state
\cite{brona95,onwoa95}. 

However, the success of simple lattice models also states their limits. 
 
The results obtained so far constitute an increasing evidence of the fact that
proteins are very peculiar heteropolymers \cite{shfaa91,pagra96}: far
from  being random, they
present an "energy-landscape" with correlated minima, and a native state which
has probably been selected to be the ground state of the highest number 
of sequences (to face the risk of mutations) and to be kinetically accessible
from all  initial configurations of the chain.
The role played by local contacs along the chain (the secondary
structure), 
which had been disregarded by REM, is now coming again strongly to
attention, as it seems
to be crucial for rapid folding and resistance against mutations
\cite{lihea96}.

Simple heteropolymer models on a lattice are not well suited to deal 
with this kind of features,  and
the only way to improve our knowledge  
within this scheme would be  that of making simulations with longer 
chains: this is clearly unfeasible, as the only way to find the native
state relies on a complete enumerations of all compact configurations 
of the polymer, whose number grows exponentially with the number of residues. 

The need for an exhaustive enumeration comes from the fact that chain
connectedness is responsible for a strong frustration of a generic 
polymer, so that the energy landscape is rugged, and states of the same energy
(but very different in shape) may be found anywhere, asking for a complete
searching of the configuration space.

Real proteins circumvent this problem when folding
to the native state,
since they are provided with a sequence which, given the geometrical 
constraints of connectedness and microscopic steric hindrance 
to be fulfilled, encodes
the smallest frustration, and the smoothest  energy landscape (essentially, 
this is the statement of the "principle of minimal frustration"
\cite{brwo87,brona95})

The problem with lattice models comes from the fact that we do not
know {\sl a-priori}, given the constraints of lattice geometry (intrinsically
different from the natural ones), which 
sequences of what hydrophobic charges correspond to
the smoothest landscapes, and the only way to find it out consists in a
exhaustive numerical analysis of the entire configuration space, since no
definite hints can be provided by real proteins.

The situation would greatly improve with a model directly
related to the real systems, such that a mapping would exist between
protein and model configurations.
In this case, a direct comparison of the ground state with true native 
one would be possible, 
and one could check the goodness of the model by direct inspection.

In this paper, we present a model which, in a coarse-grained way, allows us
to deal with any chosen sequence of any lenght.
Such model  is based on a description of the protein chain in terms
of pieces of helices,
implying that the building blocks are indeed the  elements of the
protein secondary structure, which have an
"internal energy" related to Ramachandran's maps and mutually interact
according to the mean "charge" they contain.
The hamiltonian  we obtain is realistic, yet quite complicated, just
because of its generality.
However, simplified models can be extracted from it and studied
independently, and the results can be compared to real native states.

The paper is organized as follows: in Sec.~\ref{sec-mod} we define the model,
discussing the various terms in the hamiltonian, in Sec.~\ref{sec-simpmod} 
we derive
a simplified model and
calculate its ground state and partial partition function;
in Sec.~\ref{sec-conc}, we briefly summarize and comment our results.

\section{The model}
\label{sec-mod}
\subsection{Preliminary remarks}
\label{sec-prel}
We start from the observation that accurate studies of the
phenomenology reveal a number of common features of the native states
of  the majority of simple, single--domain proteins:
\begin{itemize}
\item{}{the native state is organized hierarchically in secondary,
supersecondary, and tertiary structures. Typical elements of the secondary
structure are $\alpha$--helices, $\beta$--strands and tight--turns;
supersecondary rules tell us how these elements pack together locally
(prescribing, for instance, the right-handedness of $\beta$--X--$\beta$
units), while the tertiary structure refers to the way the above mentioned
elements are arranged in space. As a general remark, one can say that 
``pieces of secondary structure that are adjacent in the sequence are also
often in contact in three dimensions'' \cite{chot84}; knots in the
chain seem also to be generally forbidden };
\item{}{the native state is highly compact, with the non--polar
residues buried on the inside, in order to minimize their contact with
the solvent. The urge to protect the hydrophobic residues from water
is believed to be the leading factor in the folding process: the
secondary and supersecondary structures would emerge in order to
accomodate in the best way the hydrophobic core, with the minimal
frustration of local interactions;}
\item{}{the periodicity of the helices tends to mimick that of the
sequence, when there is one. It is known, for instance,  that
$\alpha$--helices on the surface of the protein's native state
usually present an external side, exposed to the solvent, with polar
residues,  while the other one is hydrophobic;}
\item{}{the partial success of structure identification methods, based
on the analysis of  the homologies between sequences, suggests that,
even though the folding process is dominated by hydrophobic
interactions, local properties pose serious constraints on the final
structure, and somewhat limit the number of possible choices.}
\end{itemize}

For the above reasons, we aim to construct  a model able to handle both the
local and the global aspects of the main chain geometry (disregarding 
side-chain configurations).

One way to cope with the two contrasting requirements of a
coarse--grained ``effective'' modelling of the interactions, and of a
good control of the local constraints imposed by  steric hindrance
and chain connectedness, is to
think  of the protein as made up by pieces of
different helices, linked together one after the other. 
This picture is general enough to describe probably all the relevant
conformations of a protein: it is built having in mind the above-mentioned
features of native states, but it may also represent chains in
coil conformations, when many small helices are present, with
random orientation. For the consistency of this approach, we shall
approximate with a helix a part of the protein at
least three peptide units long; shorter helices will not be allowed.
Since a perfect helix involves  repetition of a fixed 
dihedral angle ($\phi,\psi$) at each peptide unit, 
this approximation may  appear to be somewhat  crude when there is a
strong local 
variability of the above variables, as it happens in loops and turns.
However, this is not a major problem, because a tight turn can be
fairly well represented by a short piece of regular helix, and the
shortness
implies that only a small error  in the
energy is introduced. The same holds true for any ``coil'' region of
the chain, which may be partitioned and treated in the
same way.

The helices are described by their radius and pitch, their lenght, and
the orientation of a reference frame attached to each of them, which
specifies the direction of their axis. We shall see later that these
are not the most useful variables to introduce in the hamiltonian, but
we start with them for the sake of simplicity.

The equation of the curve representing the protein chain is assumed to be:
\begin{equation}
{\bf r}(s) \,=\, \sum_{i=1}^{N_h} b_i(s)\,{\bf h}_i(s) \;\;\; ,
\label{chain}
\end{equation}
where $s$ is a continuous variable parametrizing the curve points, and
ranging from $0$ to $N$, the total number of residues; $b_i(s)$ is the 
limit for $\lambda \rightarrow 0$ of the function 
\[ b_i(s,\lambda) = -g_i(s,\lambda) + g_{i-1}(s,\lambda) \;\;\;,\] 
which represents a ``barrier'': $g_i(s,\lambda)$ is a function of the variable
$s$ designed in such a way that in the
limit  $\lambda \rightarrow 0$ it becomes a step function. For instance
one could choose
\[ g_i(s,\lambda) = \frac{1}{2} \tanh(\frac{s-s_i}{\lambda})  \;\;\;, \]
whereby $b_i(s_{i-1},0) = b_i(s_i,0)=1/2$, and 
\[\dot{g_i}(s,0) \equiv \frac{dg_i}{ds}(s,0)=\delta(s-s_i) \;\;\; . \]

The ${\bf h}_i$ are the helices expressed in their reference frame
(${\bf e}_{1,i},{\bf e}_{2,i},{\bf e}_{3,i}$):
\begin{eqnarray}
{\bf h}_i(s) &=& a_i \,\left[\, \left( \cos(u_i(s-s_{i-1}))-1
\right)\, {\bf e}_{1,i} \,+ 
\sin(u_i(s-s_{i-1})) \,{\bf e}_{2,i} \,+ \right. \nonumber \\ 
&& \left.  u_i h_i(s-s_{i-1})\,{\bf e}_{3,i}\,\right] + 
{\bf h}_{i-1}(s_{i-1}) \;\;\;,
\label{hel}
\end{eqnarray}
labeled so that helix i starts at $s_{i-1}$ and ends at
$s_i$, with $s_0=0$ and $s_{N_h}= N$. $N_h$ is the total number of
helices, residues are labeled from 1 to $N$, and the convention holds
that a residue sitting at the junction between two helices belongs to
the first of them. We will name from now on
$n_i = s_i - s_{i-1}$ the lenght of  helix $i$.
We choose
\be
u_i = \sigma_i \frac{L}{a_i \sqrt{1+h_i^2}} \;\;\;,
\label{u}
\ee
where L is the lenght of a peptide unit (the distance between two neighboring 
$\alpha$-carbon atoms),
so that  the line element on each helix is
$\left|\dot{{\bf h}}_i\right| ds = L ds$. We assume the sign
$\sigma_i = \pm 1$ of $u_i$
positive for right-handed and negative for left-handed helices,
while the product $u_i h_i$ is always positive.
We also ask that helices have the same lenght of the piece of chain they
represent: this may be done by requiring that $\Delta s=1$ when we move 
along the protein chain of one peptide unit: in this way the above
defined $n_i$ coincides with the number of residues in the secondary
structure element that the helix describes.

We see that six scalar variables are needed to specify a helix: they are
$n_i$, $a_i$, $h_i$, and three Euler rotation angles 
relating the helix reference frame to the fixed "laboratory" one.
For an infinite helix, the radius $a_i$, the pitch $h_i$ and the angular
parameter $u_i$ are 
related to curvature $\kappa_i$ and torsion $\tau_i$ in a straightforward way:
\begin{equation}
a_i = \frac{\kappa_i}{\kappa_i^2 + \tau_i^2} \;\;,\;\;
h_i = \frac{\tau_i}{\kappa_i}\;\;,\;\; 
u_i = L \sigma_i (\kappa_i^2 + \tau_i^2)^\frac{1}{2} \;\;\;.
\label{ahu}
\end{equation}

Obviously curvature and torsion are the natural candidates to appear
in the local part of the hamiltonian, which will take into account the 
stiffness of the chain and the steric hindrance of the residues. One is 
therefore led to study the form of curvature and torsion of the 
curve in Eq.(\ref{chain}). 
After some lengthy but straightforward manipulations, remembering that
${\bf h}_i(s_i)={\bf h}_{i+1}(s_i)$
and taking into account only the leading terms in the
limit  $\lambda \rightarrow 0$, one finds that these quantities have
the form of a sum involving
curvatures and torsions of the various helices in the chain,
plus some terms coming  from the regions near each junction, depending only
on the two neighboring helices. 

For instance we obtain for the curvature (see Appendix~\ref{app1}):
\begin{equation}
\kappa = \sum_{i=1}^{N_h} \left( \kappa_i \vartheta(s-s_{i-1}) \vartheta(s_i-s)
+   \delta(s-s_i) \left( 8 \frac{1- (\dot{{\bf h}}_i \cdot 
\dot{{\bf h}}_{i+1})^2 }
{(1+  \dot{{\bf h}}_i \cdot \dot{{\bf h}}_{i+1})^3} \right)^\frac{1}{2} \right)
\label{chaink}
\end{equation}
where dots indicate derivatives with respect to $s$, and $\delta(s-s_i)
\; , \; \vartheta(s-s_i)$ are respectively Dirac delta and Heaviside
theta functions.

We see that the relevant quantity at the interface is the scalar product 
$\dot{{\bf h}}_i \cdot \dot{{\bf h}}_{i+1}$ 
between the right and the left limit in $s = s_i$ of the tangent
vectors.
Similar results hold for the torsion, as well as for any other
quantity  obtained
by algebraic operations on curvature and torsion.

The fact that it is possible to reduce the expressions of curvature and 
torsion of the whole chain 
to sums of the corresponding quantities for each helix, plus "interface terms"
depending only on nearby elements, suggests  
that also the hamiltonian may be built as 
a sum of "local" single-helix terms with next-neighbour
interactions, accounting for the stiffness of the
chain. In addition to these, a third, non-local term, will describe 
the interactions
between non-neighbouring helices. 
Therefore we write:
\[
H = \sum_{i=1}^{N_h} (H_i + H_{i,i+1}) + \sum_{i<j=2}^{N_h} H_{i,j}
\;\;\;. 
\]

The protein sequence will come into play in the last term, because the
interaction between helices obviously depends on the residues they are made of,
but will also have a role in the first one, as helices are
preferred if they present the same local periodicity as the sequence.
 
The explicit form of the hamiltonian will be
discussed in section~\ref{secham}; in the next one, 
we introduce a formalism allowing us  to treat
conveniently the non-local term, which is awkward to handle in the
variables that appear in Eqs. (\ref{chain}) and (\ref{hel}).

\subsection{Dynamical variables}
\label{sec-dynvar}
In order to specify the position and the kind of each helix, we
introduce  the following variables:

\be
\begin{array}{ccll}
N_h\!\!\!&&\!\!\hbox{the total number of helices} \;\;\;& \\
n_i\!\!\!&=&\!\!s_i - s_{i-1} &(n_i \in [p_1,p_2])\\
l_i\!\!\!\!&=&\!\!\frac{1}{2}\left(s_i + s_{i-1}+1 \right) & 
(l_i \in [q_{1,i},q_{2,i}])\\
{\bf v}_i\!\!\!&=&\!\!{\bf h}_i(s_i) - {\bf h}_i(s_{i-1})& \\
{\bf B}_i\!\!\!&=&\!\!\frac{1}{2}({\bf h}_i(s_i) + {\bf h}_i(s_{i-1}))&  
\end{array}
\label{dyva}
\ee
where $p_2 = N - (N_h-1) p_1$ and $p_1=3$, since a helix cannot be
defined with less than three residues; 
$q_{1,i} = \frac{1}{2}[1+p_1 (2 i-1)]$, $ q_{2,i}=
N+\frac{1}{2}[1 - p_1 (2 (N_h-i)+1)]$.
In the above equations $n_i$ is
the lenght of the $i$-th helix  expressed in residues; $i \in [1, N_h]$;
 $l_i$ represents
the position along the sequence of the center of the $i$-th helix;
${\bf v}_i$ is
the vector joining the end-points of helix ${\bf h}_i$; it is  the 
geometrical analogue of $n_i$; ${\bf B}_i$ is the
the spatial position of the middle point of ${\bf v}_i$.

Two other variables, related to curvature and torsion, are needed to
completely 
specify the characteristics of a helix: a useful choice, which will allow
us to
write a realistic potential in a simple form (see Section~\ref{sec-hi}
below), is to
introduce:
\bq
z_i &=& \frac{L \tau_i}{u_i} \;\;\;, \\
\label{z}
w_i &=& u_i - 2 \pi \vartheta(-u_i)\;\;\;,
\label{w}
\eq
where $u_i$ has been defined  in Eq. (\ref{ahu}). Note that $z_i$ ranges
between 0 and 1, while $w_i \in [0, 2\pi]$. 
The definition of $w_i$ allows us to remove the discontinuity between
right and
left-handed helices at $u=\pm \pi$, which is model-induced but
inevitable in a
description of the chain in term of helices. A chain in such a
conformation, where there are
exactly two residues per turn, may be regarded as both  left-handed
and  right-handed, and a little deformation can bring to the
one or the other type of configuration; yet there is no  continuous
operation  that
trasforms a right-handed helix into a left-handed one. This is
reflected in the fact that one abruptly passes from $u=\pi$ to $u=-\pi$
while smoothly  deforming the chain.

The above variables specify how the chain is partitioned into helices
and is embedded in three-dimensional space, providing a
sequence-independent formalism.
The sequence enters the model through new variables $q_k$ ($k=1 \ldots N$) 
and ${\bf p}_{\perp}^2(l,w)$.
The former are related to the nature of each residue $k$, and measure  
its coupling to the other residues. In the following we shall refer to 
Li and coworkers~\cite{litaa95} who write the Mijazawa-Jernigan  
interaction matrix \cite{mije85} as:
\be
M_{\rho \sigma} = \mu_0 + \mu_1 (q_\rho + q_\sigma) + \mu_2 q_\rho q_\sigma
\;\;\;\;\;\;\;\; (\rho,\sigma=1 \ldots 20)
\label{mijemx}
\ee
(a slight change of notation is performed here with respect to
the original paper). This equation can be recast as:
\be
M_{\rho \sigma} = Q_\rho + Q_\sigma + \frac{\mu_2}{2} (q_\rho - q_\sigma)^2
\label{mijemx2}
\ee
with
\be
Q_\rho=\mu_0/2 + \mu_1 q_\rho  + (\mu_2/2) q_\rho^2\;\;\;.
\label{Q}
\ee
The authors show that $Q_\rho$ correlates well with the hydrophobicity of
the residues; for this reason we can call it "the hydrophobic charge"
of aminoacid $\rho$.

Since we deal with entire helices at a time, and not with single
residues, we shall introduce the average $q$ of a helix, centered
in $l_i=l$, as
\be
\overline{q}(l) = \cases{\frac{1}{2m + 1}\sum_{j=-m}^m q_{l+j},&if $l=1,2,\ldots$ \cr
\frac{1}{2(2m + 1)}\sum_{j=-m}^m (q_{l-\frac{1}{2}+j}+q_{l+\frac{1}{2}+j}),
&if $l=\frac{1}{2},\frac{3}{2},\ldots$}
\label{qbar}
\ee
(integer or half-integer values of $l$ are the only ones allowed for 
the central points of the helices, $l_i$;
the variable $m$ is an arbitrary number, 
comparable with the mean lenght of the helices).

The corresponding values for the hydrophobic charge $\overline{Q}(l)$
are obtained from Eq.~(\ref{Q}) with  $q$ replaced by $\overline{q}$.
Notice that this is the correct way of evaluating 
$\overline{Q}(l)$, since it is easy to show that the
average interaction between $n_1$ residues on a helix and $n_2$ on
another is given by:
\bqn
\overline{M}&=&\frac 1{n_1n_2}\sum_{i=1}^{n_1}\sum_{j=1}^{n_2}M_{ij}=
\frac 1{n_1n_2}\left[ n_1n_2\mu _0+\mu _1\left( n_2\sum_{i=1}^{n_1}q_i
+n_1\sum_{j=1}^{n_2}q_j\right) +
\mu _2\sum_{i=1}^{n_1}\sum_{j=1}^{n_2}q_iq_j\right] =\\
&=& \mu _0+\mu _1\left( \overline{q}_1+\overline{q}_2\right) +
\mu _2\overline{q}_1\overline{q}_2 \;\;\;,
\eqn
and hence is naturally written as a function of the average 
$\overline{q}_1$ and $\overline{q}_2$ on the helices.\\

The other variables are related to the local periodicity of the
sequence, and are defined in the following way. Considering a generic
helix of lenght $2n +1$, 
centered at the point $l$ along
the sequence, the quantity:
\[{\bf p}_{\perp }(l,w,n) 
= \frac{1}{\sum_{k=-n}^n Q_{l+k}} \sum_{k=-n}^{n} Q_{l+k} 
\left(\cos[w (l+k)] {\bf e}_{1} + \sin[w (l+k)] {\bf e}_{2}  \right)\] 
is the projection, on the plane perpendicular to the helix axis,
of the "hydrophobic dipole moment" calculated at a point on the axis,
and normalized with respect to the total charge $ \sum_{j=-n}^n Q_{l+j}$
(for the sake of simplicity, we take $l$ to be an integer in these equations).

We observe that ${\bf p}_{\perp }^2$ reveals the 
prevalence of non polar residues on one side of the helix, characterized
by the periodicity $w$.
Therefore, 
in Sec.~\ref{sec-hi} we shall write a local hamiltonian depending explicitly on
\be
{\bf p}_{\perp}^2(l,w,n) = \frac{1}{(\sum_{j=-n}^n Q_{l+j})^2}
 \sum_{j,k=-n}^{n} Q_{l+j} Q_{l+k}  \cos((j-k) w) \;\;\;,
\label{pquad}
\ee 
and favouring configurations which maximize ${\bf p}_{\perp}^2$.

The dinamical variables previously defined are not completely independent
from each other, and the
following constraints hold:
\begin{enumerate}
\item{the sum of the residues of all the helices must be equal to the
total lenght of the chain:
\be
\sum_{i=1}^{N_h} n_i - N =0
\label{vinc0}
\ee
}
\item{the lenght of the end-to-end vector $v_i$ is related to the
lenght and shape of the helix:
\be
{\bf v}_i^2 - \left| {\bf h}_i(s_i) -
{\bf h}_i(s_{i-1}) \right|^2  
\equiv  {\bf v}_i^2 - n_i^2 L^2 
\left[ z_i^2 + (1-z_i^2)
\frac{\sin^2(\theta_i)}{\theta_i^2}\right] = 0 \;\;\;, 
\label{vinc1}
\ee
where $\theta_i = n_i u_i / 2$;}
\item{the end of one helix must coincide with the beginning of the
following one, both in sequence and in space:
\bq
 {\bf B}_i - {\bf B}_{i-1} - \frac{({\bf v}_i + {\bf
v}_{i-1})}{2}&=& 0\;\;\;,
\label{vinc2} \\
l_i - l_{i-1} -
\frac{n_i + n_{i-1}}{2} \;\;&=& 0\;\;\;.
\label{vinc3}
\eq   
}
\end{enumerate}
In these equations, $i$ ranges from 1 to $N_h$, and, to be consistent
with the definitions of $l_i$, we set $l_0 = 1/2$, $n_0 = 0$. 

From the above discussion the following picture emerges: we describe the
geometric shape of the protein by the variables  ${\bf B}_i,{\bf v}_i,
w_i, z_i, n_i, l_i$; besides, we  give the
sequence of 
``charges'' $q_k$ which represent its residues. 

Notice that the above variables do not specify completely the position of
the helices in space, because, when the constrains are satisfied,
there is still a complete degeneracy for rotations of  each helix 
${\bf h}_i$ around the vector ${\bf v}_i$.

Actually, given a chain conformation, only one of these degenerate 
configurations, which are
indistinguishable in our scheme, corresponds to it.
Therefore, provided  we choose a good criterion to identify
helices out of real chain conformations, we have that any protein 
configuration can be mapped in exactly one model configuration.

We may ask if the converse is also true: considering a given a model
helix, specified by the variables $n$,$w$,$z$,${\bf v}$, with
its end points pinned at $P_1 = {\bf B - v}/2$, $P_2 = {\bf B + v}/2$,
we see that:
\begin{itemize}
\item[a.]{many chain conformations, slightly
different from one another, correspond to it, since real helices
are not made up by an exact repetition of dihedral angles; anyway, we
assume we can consider these to be represented by the same geometrical
helix;}
\item[b.]{it's quite unlikely that  two (or more) real helices, with
shape and lenght well described by the above variables, but
corresponding to different degenerate positions around vector ${\bf
v}$, may exist;}
\item[c.]{it could be possible, on the other hand, that no real helix
with the above parameters could fit between $P_1$ and $P_2$, due to
the stiffness of the chain at those points.}
\end{itemize}
If case 'c' holds, troubles arise, since forbidden chain configurations
could appear as allowed model ones.

In the following, we shall make the assumption that, given a model
helix  (of lenght $n \ge 3$), with its end-points at $P_1$ and $P_2$, it is
always possible to replace it with the corresponding real chain helix,
in such a way that no relevant perturbation is introduced in the total
energy of the protein.

This is reasonable, since small adjustements of the chain, out of a
perfectly regular configuration (yet not affecting its overall helical
shape), can intervene and prevent forbidden joint conformations, thus
reducing the energetic penalty to a small fraction of the total energy.
We shall come back to the discussion of the energy contributions from
the helix  junctions in the next section, where we discuss the 
explicit form of the hamiltonian.


\subsection{The Hamiltonian}
\label{secham}

\subsubsection{The local hamiltonians $H_i$}
\label{sec-hi}
Two terms will contribute to the internal energy of a helix: the first
one, $H_i^{0}$, 
is purely geometric, and takes into account the experimental
Ramachandran plot to dictate which kind of helices are more likely to be
formed.
The second one, $H_i^{1}$, on which we commented above, is sequence   
dependent, and acts as an
external field, biasing the helices towards a certain periodicity.

It is important to notice that, if proteins were made  by a
sequence  of exact repetitions of the same $(\phi,\psi)$ angles, the
association of a geometrical helix to each part of the chain would be
straightforward and the geometrical quantities $w,z$ of the  helix
could be written as functions of $(\phi,\psi)$:
\be
\left\{ \begin{array}{rcl} 
  \cos(\frac{w}{2}) &=& \sigma(s_2) c(\phi,\psi) \;\;\;, \\
  z &=& \frac{\left| s_2 \right|}{\sqrt{1 - c^2(\phi,\psi)}}\;\;\;,  
\end{array} \right.
\label{inv}
\ee
where we have defined:
\[
\sigma(x) = \cases{+1,&if $x > 0$ \cr
                     -1,&if $x \le 0$}\;\;\;,
\]
and use the explicit expressions 
\bqn
c(\phi,\psi) &=& a \sin(\frac{\psi + \phi}{2}) + b \sin(\frac{\psi -
\phi}{2}) \;\;\;,\\
s_2 &=& c \cos(\frac{\psi + \phi}{2}) + d \cos(\frac{\psi - \phi}{2})\;\;\;,
\eqn
(see Appendix~\ref{app2} for details). The mapping $(\phi,\psi)  \rightarrow
(w, z)$ in the above equations is two-to-one, as can be seen 
from the study of the solutions of
the fourth degree polynomial equations involved in the inversion of
Eqs.(\ref{inv}). This reflects the fact that $(\phi,\psi)$ provide a
complete description of the geometry, specifying not only the position
of the $C_\alpha$ atoms, but also the orientation of the peptide
planes, which have been disregarded in our approach.

Dealing with real proteins implies that each helix is associated to
a portion of chain  where a certain amount of irregularity in $(\phi,\psi)$
is inevitable. This has no practical consequences when the irregulaties
are small and the dihedral angles are clustered  around a particular
position (which happens for long elements of secondary structure). However,
for short coil regions with a great variability in the dihedrals, it
would be quite artificial to relate the
best fitting values of $w,z$ to an hypothetical $(\phi,\psi)$
repeating couple. 
It is clear that in the latter case the relationship
between $(w,z)$ and $(\phi,\psi)$ is weakened.
For this reason we shall make the simplifying assumption that  $w, z$
can be assumed as fundamental variables, and that the helix-model 
configurations are in a one-to-one
relationship with them. 
Since $(w, z)$-couples that cannot correspond to any real
protein configuration are introduced by this ansatz, we shall write 
the hamiltonian $H_i^0$ in
such a way that these values have a vanishing weight in the
evaluation of the partition function.

The choice of an  explicit expression for the hamiltonian requires a
careful analysis, because no reliable potential function based on
first principles is known. This is not surprising, if one thinks that
each residue is itself a many-body system, usually in interaction with
the solvent molecules. 
On the other hand, the fact that even a crude
hard-sphere model for a dipeptide reproduces the experimental 
Ramachandran's maps in an essentially correct way (compare for
instance Fig.~(12A, 13A) in Ramachandran's
article~\cite{rasa68} with Fig.~(5) in Morris {\it et al.}~\cite{momaa92}),
implies that a
simplified description of the potential function should be possible,
and suggests its main characteristics.

We already mentioned the fact that, even in the case a perfect helical-shaped
chain, the variables $w$, $z$ give a description of the chain geometry
which is 
less detailed than that provided by $(\phi,\psi)$
angles, and even more so in comparison with an all-atom description.  
Despite this, it is possible to write  a
potential function in the variables $w$, $z$ which correctly
reproduces the main features of the Ramachandran's plots, in the form
\be
H_i^0 = (n_i-1) \gamma_0
     \left[c_1 \left((w_i- c_2)^2 - c_3 \right)^2 + c_4 
     +c_5 \left(z_i-c_6 + c_7 (w_i-c_8)^2 \right)^2 \right]\;\;\;. 
\label{hi0}
\ee
Here the $c_k$'s are fixed adimensional constants; 
the factor $(n_i-1)$ takes into account the feature that each residue 
in the helix
feels the same potential, except the last one, which corresponds to the
junction with the following helix, and must be treated in a different way.
In Figs.(\ref{figv4wz},\ref{figv4fipsi}) we  plotted the contour lines of
this potential, with 
$n_i -1 =1$, respectively as a
function of  
$w$, $z$, and of their images on the    $(\phi,\psi)$ plane. 

Note that, since $b/a \approx 1/20$, there is nearly 
symmetry under the exchange of $\phi$ and $\psi$, so that there is an
intrinsic difficulty in distinguishing the region above from that
below the diagonal $\phi=\psi$ by means of the quantities at the right
member of Eq.(\ref{inv}). 
This can be considered a minor problem, since the physically allowed region
roughly coincides with that above the diagonal $\phi=\psi$
and  we aim to study,
as a first approach, only the behaviour of a system which presents two energy 
minima roughly corresponding to the $\alpha$- and $\beta$-regions of
Ramachandran's 
map (as a consequence, we disregard left-handed $\alpha$-helices, since a    
residue in 
that position usually belongs to a turn, and not to an actual left-handed 
helix). \\

The above {\sl ad hoc} hamiltonian provides
a correct qualitative description of the phenomenological results,
without introducing expressions more complicated than a fourth degree
polynomial. In the following we shall consider it as an unperturbed
hamiltonian, to which the non local-interactions add as 
perturbations, which remove its degeneracy without affecting its
overall shape in plane $(\phi,\psi)$.

Coming to the sequence dependent part of the local hamiltonian,
a very natural choice is that of taking 
\be
H_i^1 = -\gamma_1  n_i P(l_i,w_i)\;\;\;, 
\label{hi1}
\ee
where $\gamma_1$ is an appropriate dimensional
constant, and 
$P(l_i,w_i) = {\cal F}({\bf p}_{\perp}^2(l_i,w_i,n))$
is some simple function of ${\bf p}_{\perp}^2(l_i,w_i,n)$. For
instance  ${\cal F}$     
could be either the  average of ${\bf p}_{\perp}^2$ calculated for 
different accessible lenghts of the helices (the values of $n$ in 
Eq.~(\ref{pquad})), or ${\bf p}_{\perp}^2$ itself, 
evaluated with a particular phenomenological mean value of $n$.

A more detailed study on the best expression for
$\cal F$ is left to future work on the subject: in the following,
we shall choose a particular form for $\cal F$ only when,   
in section \ref{sec-simpmod}, we shall study the ground state of
hamiltonian 
Eq.(\ref{hi1}) for
a small synthetic protein,
showing that $P(l_i,w_i)$ can indeed provide partial
information about the native state.  
   

\subsubsection{The interaction between neighbouring helices $H_{i,i+1}$}
\label{sec-hnn}
So far we have disregarded the contributions to the energy coming from
the aminoacids at the junctions between helices: they have been kept
out from Eq.(\ref{hi0}) thanks to the factor $(n_i-1)$.  When we try
to keep them into account, we immediately face many difficulties,
since there is apparently no natural way to relate exactly the
microscopic potential determining the possible $(\phi,\psi)$ values to
the description of the chain in terms of helices.

We saw in Eq.(\ref{chaink}) that, at the junctions between helices,
the curvature depends on the scalar product of the right and left
limits of tangent vectors in those points.  This would suggest to
write an interaction penalizing discontinuities in the tangent vector;
yet, this is quite awkward for the following reasons:
\begin{itemize}
\item{while it is possible to relate curvature and torsion of a helix to
the dihedral angles specifing the chain, it is extremely complex to do
the same for the scalar product $\dot{{\bf h}}_i \cdot \dot{{\bf
h}}_{i+1}$: the knowledge of $(\phi,\psi)$ at the the junction is not
sufficient to specify the direction of the tangent vectors (remember
that the $C_{\alpha}$ atoms of the chain do not even lie on the
helices, due to the requirement that lenghts be the same when measured
along the (continuous) helices or the (discrete) chain);}
\item{even if it were possible to find a mapping, relating the dihedral
angles at the junction to the helix geometrical description, in terms
of tangent vectors, uncontrollable mistakes would be done in
evaluating the energy. In fact, real helix junctions are different
from ideal ones, since structural adjustments are allowed in real
chains to minimize the energetic cost of the junction which cannot be
described by ideal chain geometry, where helices are stiff.  Hence, if
we are interested in a reasonable estimate of the energy, a detailed
description of the tangent vectors' dependence on the $(\phi,\psi)$
angles at the junction is essentially useless.}
\item{Finally, tangent vectors are very difficult to write down
within the adopted formalism (because of the degeneracy under
rotations, discussed in Sec.~\ref{sec-dynvar}, even if we don't relate them to
real chain quantities.}
\end{itemize}

Therefore, we have to write down an expression for the interaction
energy between neighbouring helices without relating it to the real
chain geometry, and without resorting to tangent vectors.  The natural
candidates to appear in such an expression are, of course, the vectors
${\bf v}_i$, ${\bf v}_{i+1}$, but the functional form to be chosen is
by no means obvious.

In principle one could study first the total energy of two successive
secondary structure elements (for instance, by looking at dihedral
angles and applying microscopic potentials) as it comes out from
phenomenology, then represent such elements with model helices, and
finally get the interaction energy as the difference between the total
and the sum of the single helices'.  It would be possible, in this
way, to relate the interaction energy to the relative positions of
${\bf v}_i$ and ${\bf v}_{i+1}$.

Yet, as a first approach, we assume this energy to be a costant, at
each junction, regardless of the values of the dynamical variables:
for instance, we can take the mean energy of the residues calculated
with the Ramachandran's map distribution.  Since $N_h-1$ is the
number of junctions, we set:
\be
H_{nn} = \gamma_2 (N_h-1)\;\;\;.
\label{hnn}
\ee
This simple hypothesis entails an important effect: if one neglects
the sequence--dependent hamiltonian $H_i^1$ and the non local
interactions, splitting up a helix in two pieces with the same $(w,z)$
as the former one involves the substitution of a residue of energy
$H_i^0(w,z)$ with a residue of mean energy $\gamma_2$.  Hence helix
breaking will be penalized for helices with "good" values of $(w,z)$
($\alpha$-helices and $\beta$-strands, for instance), and favoured in
the opposite case.
 
This is very important, because both entropic effects and non-local
interactions, as we shall see below, would favour configurations with
many short helices, regardless of $(w,z)$ values: $H_{nn}$ competes
with the above effects, allowing, in principle, the existence of
equilibrium states of the model presenting long elements of secondary
structure.


\subsubsection{ The non-local interactions $H_{ij}$}

The modeling of non-local interactions requires a careful analysis of
their nature and characteristics.  Two different contributions are to
be dealt with: hydrogen bonds and hydrophobic interactions. The former
are responsible of orientational preferences of the couplings between
elements of the secondary structure, but are believed to play an
insignificant role as a driving force for folding, since hydrogen
bonds to solvent are of the same energy than intramolecular ones.  The
latter are responsible for the collapse of the chain to globular
states, but are very difficult to model, since they come mostly from
entropic effects invoving the solvent, and not from a true coupling
between residues.  For these reasons, we abandon the idea of writing
non-local interactions on the grounds of microscopic considerations,
and once more resort to phenomenology.
 
First of all we notice that typical distances between the axes of
interacting secondary structure elements in the native state range
from 0.46 nm (two hydrogen-bonded $\beta$-strands\cite{grkh94}) 
to about 1 nm (two $\alpha$-helices or two
$\beta$-sheets, \cite{chot84}).  Therefore we simply write down an
attractive square-well potential in the variable $\Delta B_{ij} =
\left| {\bf B}_i- {\bf B}_j \right|$, taking ${\bf B}_i$ as
reprentative of the $i$-th helix, with a hard core repulsion
preventing overlap between helices.

Then, we look at the phenomenology of non-local interactions in the
native state, considering at first only the hydrophobic effect and
disregarding hydrogen bonds in $\beta$-sheets.  We see that usually
two elements of secondary structure tend to pack as closely as
possible, just due to the hydrophobic effect. For geometrical reasons
this usually means that they cannot be parallel; thus the number of
residues which are actually into contact is independent of the lenght
of the helices, and also, roughly, of their characteristics.  If,
following Li and coworkers~\cite{litaa95}, 
we take as the "microscopic" contact
interaction between
two residues that given in Eq.~(\ref{mijemx}),
we can write for the interaction between helices:
\bq
H_{ij} &=& \vartheta(\rho_1 - \Delta B_{ij}) \vartheta(\Delta B_{ij}-\rho_0) 
\left[ \gamma_3
\chi \left(\mu_0 + \mu_1 \left(\overline{q}(l_i) + \overline{q}(l_j)\right)
+ \mu_2  \overline{q}(l_i) \overline{q}(l_j) \right) \right] +  \nonumber\\
&&+\gamma_4 \vartheta(\rho_0 - \Delta B_{ij})\;,
\label{hij}
\eq 
where $\overline{q}(l_i)$ are the quantities defined in Eq.(\ref{qbar});
$\chi$ is the
average number of contacts between residues in two close-packed
elements of the secondary structure, $\gamma_4 \gg 0$ provides an
hard-core repulsion when the distance is less than $\rho_0$; $\rho_1$ is the
range of the attractive interaction 
$\gamma_3 >
0$ "normalizes" the interaction with respect to the other terms in the
hamiltonian: again, it should be small compared to $\gamma_0$.

Coming to the hydrogen bonds in a $\beta$-sheet, we see that they tend
to align the two interacting strands, independently of their charge.
Yet, it is known that $\beta$-sheets show very little stability when
exposed to the solvent, because residues easily form hydrogen bonds
with water. In our approach, where the solvent is taken into account
implicitly in the coupling strenght, one should relate
hydrogen-mediated interactions to the geometry of the helix and to the
surrounding environment (in order to distinguish between exposed and
buried sheets). Hence, hydrogen bonds between $\beta$-strands should
depend on the overall hydrophobic charge of the environment they are
embedded in: this is far too complex to be described exactly.

For the sake of simplicity, we shall use Eq.(\ref{hij}) also to
describe interaction between $\beta$-strands, neglecting the tendency
towards alignment.  A more detailed representation would involve the
introduction of terms involving ${\bf v}_i \cdot {\bf v}_j$, and also,
perhaps, of ${\bf v}_{i-1} \wedge {\bf v}_{i+1} \cdot {\bf v}_i$, to
account for right-handedness of super-secondary structures like
$\beta$-X-$\beta$.
\\

As a result of the above discussion, the complete hamiltonian of our
model, also including the constraints, reads:
\be
H = H_{nn} + \sum_{i=1}^{N_h} (H_i^0 + H_i^1) + \sum_{i<j=2}^{N_h}
H_{i,j} + {\cal V}^0 + \sum_{i=1}^{N_h}({\cal V}_i^1 + {\cal
V}_{i-1,i}^2 + {\cal V}_{i-1,i}^3) \;\;,
\label{protham}
\ee
where, recalling here all the results for the sake of clearness:
\bq
H_{nn} &=& \gamma_2 (N_h-1) \;\;\;,\nonumber \\ H_i^0 &=& (n_i-1)
\gamma_0 \left[c_1 \left((w_i- c_2)^2 - c_3 \right)^2 + c_4 +c_5
\left(z_i-c_6 + c_7 (w_i-c_8)^2 \right)^2 \right] \;\;, \nonumber \\
H_i^1 &=& - \gamma_1 n_i P(l_i,w_i)\;\;\;,\nonumber \\
H_{ij} &=& \vartheta(\rho_1 - \Delta B_{ij}) \vartheta(\Delta B_{ij}-\rho_0) 
\left[ \gamma_3
\chi \left(\mu_0 + \mu_1 \left(\overline{q}(l_i) + \overline{q}(l_j)\right)
+ \mu_2  \overline{q}(l_i) \overline{q}(l_j) \right) \right] + \nonumber\\ 
&&+ \gamma_4 \vartheta(\rho_0- \Delta B_{ij})\;,\nonumber\\
{\cal V}^0 &=& \lambda_0 (\sum_{i=1}^{N_h} n_i - N) \;\;\;,\nonumber
\\ {\cal V}_{i}^1 &=&
\lambda_{1,i} \left({\bf v}_i^2 - n_i^2 L^2 z_i^2
         \right) \;\;\;,\nonumber \\ {\cal V}_{i-1,i}^2 &=&
\mbox{\boldmath $\lambda_{2,i}$} ({\bf B}_{i} - {\bf
B}_{i-1}-\frac{({\bf v}_{i} + {\bf v}_{i-1})}{2}) \;\;\;,\nonumber \\
{\cal V}_{i-1,i}^3 &=& \lambda_{3,i} \left( l_{i} - l_{i-1} -
\frac{n_{i} +n_{i-1}}{2} \right) \;\;\;,\nonumber
\eq
The $\lambda_{a,i}$'s are Lagrange multipliers that allow us to insert
the appropriate constraints in the hamiltonian.

Notice that we have introduced an approximated expression of ${\cal
V}_i^1$ (compare it with Eq.(\ref{vinc1})): this is possible because,
for the most significative regions of $(w,z)$-plane, with any allowed
value of $n_i$ (remember that $n_i \ge 3$), the $u$-dependent term in
Eq.(\ref{vinc1}) is negligible.

Without loss of generality, we can moreover take $c_4 = 0$: in fact,
$c_4$ can always be eliminated by the transformation:
\[\lambda_0 = \lambda'_0 - \gamma_0 c_4 \;\;\;\; , \;\;\;\;
\gamma_2 = \gamma'_2 + \gamma_0 c_4 \;\;\;,\]
whereby the hamiltonian changes of the constant term $(N-1) \gamma_0
c_4$.  The way we have chosen to implement the constraints is
particularly suitable to carry on some analytic calculations on the
model.  Obviously, it is not the only possible one, and different
choices may be useful in different approaches.
\\

An important remark concerns $H_{nn}$: the choice of an expression
independent of ${\bf v}_i$ introduces possible symmetries in the model
that could be exploited to some extent. If, in fact, we disregard the
sequence, taking $q_k=q$ as a constant and neglecting $H^1$, we see that,
given a set of $B_i$, and a choice of $n_i$, $w_i$, $z_i$, providing a
total internal energy $E^0 = \sum_{i=1}^{N_h} H^0_i$, we can certainly
change the values of $n_i$, $w_i$, $z_i$, together with the vectors
${\bf v}_i$, in such a way that $E^0$ is unchanged and the ${\bf B}_i$
fixed, so that also the non-local interaction remains the same.  This
symmetry is reduced, or removed, by the introduction of the real
charges $q_k$, but this has to be studied independently in each case.

The picture that emerges from the above hamiltonian is that of a
complicated interplay among dynamical variables: a helix of a certain
shape and lenght, specified by $w$,$z$,$n$, will be attributed an
energy based on $H^0$ (which is related to dihedral angles
conformation), plus a sequence-dependent contribution $H^1$ depending
on its position $l$ along the chain.  The values of $z$ and $n$ then
determine, through constraint ${\cal V}^1$, the lenght of the vector
${\bf v}$, and through ${\cal V}_2$, the spatial coordinates ${\bf B}$
of the helix.  The latter, in turn, determine whether the helix
considered interacts with other helices by the term $H_{i,j}$, where
the charges $\overline{q}(l)$ again depend on the internal
coordinate $l$.

For the above reasons, the hamiltonian Eq.(\ref{protham}) is inevitably
complicated. It should be remembered, though, that it describes a {\sl
generic protein}, with any sequence, and within a very realistic
framework, which deals directly with secondary structure elements.
Moreover, the model involves a reduction of the intervening number of
independent dynamical variables ($5 N_h$ against $N$ couples
$(\phi,\psi)$, if $N$ is the number of residues, with a rough estimate
for the ratio as $5 N_h/ 2 N
\approx 1/3$), so the shortest proteins could lie within the reach of
numerical studies.

In this case predictions of the model can be compared directly with
experimental findings, which could remove the need of an exhaustive search
for the ground state, that is the starting point of many lattice
models currently studied.  Quantities like:
\[\delta^2= \frac{1}{N_h} \sum_{i=1}^{N_h}
\left( \langle {\bf B}_i \rangle - {\bf B}_i^{(nat)} \right) ^2 \;\;\;,\] 
measuring the distance of equilibrium structure from the experimental
native state (the average is taken e.g. in a canonical ensemble), as
well as its analogue referring to line coordinates:
\[\sigma^2= \frac{1}{N_h} \sum_{i=1}^{N_h}
(\langle l_i \rangle - l_i^{(nat)})^2 \;\;\;,\] can give information
on the most appropriate choice of the parameters in the hamiltonian,
and on the folding transition.

Another important topic to be investigated is that of the
identification of order parameters, which could characterize the
folding transition {\it in an intrinsic way}, with no reference to a
native state known {\sl a priori}. The study of the temperature
dependence of correlation functions could possibly distinguish a true
folding transition of a ``good'' sequence from the freezing of a
``bad'' one into any minima of a rugged landscape. 

In order to get information about such thermodynamical quantities, we
must be able to evaluate the partition function associated to the
protein hamiltonian $H$. 
This is of course a very complicated task,
and  deserves a complete and specific analysis which is beyond the
scopes of the present paper.
In the next section we shall show, anyway, that one can resort
to the study of simplified models and get indeed useful information
both on the protein  under investigation, 
and on the best way  to extend the analysis 
to the more general  case of the complete model.


\section{A simplified model}
\label{sec-simpmod}
In this section, we shall mainly deal with a simplified version of the
model, where only the local terms $H_i^0 + H_i^1$ are kept into
account and the number of helices is fixed. Eventually  we shall add 
non-local interactions to it as a small perturbation, in order to retrieve
information about the spatial conformation of the protein.

When dealing with the local model, in fact, we loose 
the description of the spatial structure,
but we are left with a highly non-trivial model, with the sequence coming
into play through $H_i^1$, which provides interesting information
about both the native state and the relative importance of the interactions
stabilizing it.
Indeed, the requirement of maximizing the  separation of hydrophobic
charges on the
helices generates a scenario in which the most anphiphilic helices tend
to compete with  each other in order to grow  as long as they can.
The equilibrium configuration one finds in this way 
specifies how the
protein should be partitioned in secondary structure elements
to obtain the highest anphiphilicity. Therefore, it provides some
important insight on the secondary-structure composition of the native
state, so that it is natural to ask oneself if, at least in some cases, 
the three-dimensional
structure could  be superimposed to the resulting secondary one,
introducing non-local interactions as a small perturbation driving the
helices to the correct configuration.

In the first part of the present section we indeed show that, for the
simple synthetic  protein (already studied by Kolinski and 
coworkers \cite{kogoa93} and Raleigh
and DeGrado \cite{rade92})
specified by the sequence 
GEVEELLKKFKELWKG PRR GEIEELFKKFKELIKG PRR GEVEELLKKFKELWKG PRR
GEIEELFKKFKELIKG, the above scheme may be successfully
applied. We shall in fact find the set of $n_i$, $w_i$ corresponding
to the minimum of the local hamiltonian; then,  we shall switch on the
non-local interaction as a small perturbation  and eventually find
that native-like configurations are indeed the ground state. 
Encouraged by this result, we shall pursue the study of the local model
and, resorting to some general assumptions on  $P(l_i,w_i)$ and 
to suitable approximations, we shall find out
-- in an almost completely analytical way -- an expression for the partition
function of a generic protein, which can represent
a good starting point to study the thermodynamics
of the complete model.  

Both these investigations 
are intended as preliminary
tests, aiming to demonstrate on the one hand that the variables we
have chosen accurately describe the sequence, and on the other that
the model we propose, despite its complexity, is indeed
analytically manageable, at least in some simplified case.

We start with the hamiltonian:
\be
H(\{w_k\},\{n_k\},\{l_k\}) = \sum_{i=1}^{N_h} (H_i^0(w_i,n_i) +
H_i^1(w_i,l_i,n_i) ) \;\;\;.
\label{hsimpnum}
\ee 
where the constraints are exactly implemented, through the equations:
\be
\begin{array}{ccl}
l_i \!\!\!&=& \!\!\frac{1}{2} (n_i+1) + \sum_{k=1}^{i-1} n_k \\
n_{N_h} \!\!&=& \!\!N - \sum_{i=1}^{N_h-1} n_i
\end{array}
\label{explcstr}
\ee
We assume that the function $P(l_i,w_i)$, appearing in the
expression of $H_i^1$ (Eq.~(\ref{hi1})), has the form:
\be
P(l_i,w_i) = \cases{ p_{\perp }^2(l_i,w_i,3), &if $l_i$ is an integer \cr
\frac{1}{2}\left[p_{\perp}^2
(l_{i}-\frac{1}{2},w_i,3)+ p_{\perp}^2(l_{i}+\frac{1}{2},w_i,3)
\right], &if $l_i=k+\frac{1}{2}$, for integer $k$}
\label{bigp}
\ee
We choose $n=3$ in expression (\ref{pquad}) since this involves
calculating the hydrophobic dipole on an helix of seven residues, a
reasonable lenght both for $\alpha$-helices and for
$\beta$-strands.
We assume that $\gamma_1/\gamma_0 \ll 1$, so that we can approximate
the minima of $H$ with those of $H_i^0$, as far as $w_i$ is concerned.
In this way we are left with only two values for the $w_i$, namely
$w_{0
\alpha}$, $w_{0 \beta}$,  which in
turn entails that we can forget about $H_i^0$, whose minima are
symmetric, and only deal with $H_i^1$.
 
Since the $n_i$ are integers and the $l_i$ are integers or
half-integers, we are left with a discrete configuration space, whose
size depends on the number of helices we are considering.

Obviously, we have $2^{N_h}$ configurations for the set of $w_i$, but,
once given the set of positions $l_i$, one ``a-priori'' knows which
has the lowest energy, by a direct comparison of $P (l_i,w_{0
\alpha})$ and $P (l_i,w_{0 \beta})$ (see Fig.(\ref{figp2albe})).
Hence, we have to find the energy minimum in a space that contains as
many points as the number of possible partitions of $N$ residues in
$N_h$ helices which have a minimum lenght of $p_1$ residues.  One can
easily convince oneself that this number is given by:
\be
\pi(N_h) = \sum_{j=0}^{N_h-1} {N - N_h p_1 -1 \choose j} 
{N_h \choose j+1}\;\;\;.
\ee
For the protein considered we {\sl a priori} know that 
its native state  is made up of four $\alpha$-helices (indeed the
sequence has been designed to produce a 4-helix-bundle \cite{rade92}), 
so we try
$N_h=4$ and ask ourselves if our model will be able to find out the
correct position, lenght and kind of helices.

We set $\gamma_1=1$, $c_4=0$ and
exhaustively searched the configuration space with $N=73$ and
$N_h=4$, which contains $\pi(N_h)=41664$ points. We found that   
$(n_1,n_2,n_3,n_4)=(17,19,20,17)$, with all the helices being
$\alpha$-helices, is the ground state of our semplified model. 
This result is in excellent
agreement with the experiment, and suggests that, at
least for some proteins, the only requirement of maximal local
anphiphilicity may be enough to get the correct composition of the
secondary structure.

Encouraged by this result, we switch on 
the non-local interactions Eq.~(\ref{hij})
as a small
perturbation ($\gamma_3 \ll \gamma_1$)
to the simplified model (\ref{hsimpnum}), and 
try to predict the tertiary structure. 
We procede as follows:
relying on
the fact that the gap between the ground-state and the first ``excited
state'' of our simplified model is necessarily finite (the $n_i$
may only assume integer values), we freeze the
secondary structure $(n_1,n_2,n_3,n_4)$ we have obtained and look for
the minimum of $H_{int}=\sum_{i=1}^{N_h-1} \sum_{j=i+1}^{N_h} H_{i,j}$, where
$H_{i,j}$ is given by Eq.(\ref{hij}).
We choose in that equation the values $\rho_0=5$ \AA, $\rho_1=9$ \AA,
and show that the bundle-like
configurations are indeed the ground state (even if degenerate) of the model.
First of all we notice that, once fixed the length in residues $n_i$ of
each helix and its $z_i$ according to the fact they are all
$\alpha$-helices, we have a unique set of $v_i$, coming from Eq.~(\ref{vinc1}).
Then, we see that, upon defining ${\bf d}_i= {\bf B}_{i+1}-{\bf
B}_i$, the set of twelve cartesian components of the four vectors ${\bf
v}_i$ are subjected to the thirteen equations (see Eq.~(\ref{vinc3})):
\bq
{\bf d}_j &=& \frac{1}{2} \left( {\bf v}_j+{\bf v}_{j+1}\right)
\;\;\;\;\;\;\;\;\;\;(j=1,2,3) \\
{\bf v}_i^2 &=& v_i^2\;\;\;\;\;\;\;\;\;\;\;\;\;\;\;\;
\;\;\;\;\;\;\;(i=1,2,3,4)
\label{vincv}
\eq
One of the above equations  represents a constraint on the allowed
${\bf d}_j$: it is easy to see that Eqs.~(\ref{vincv}) lead to the
explicit expression:
\bq
v_3^2 &-& v_4^2+4d_3^2-8{\bf d}_2 {\bf d}_3+
\frac{d_3}{\left(\hat{{\bf d}}_1 \wedge \hat{{\bf
d}}_2\right)^2} \left( \hat{{\bf d}}_1\wedge \hat{{\bf d}}_2\right)
\times \nonumber\\
&&
\times \left[ \left( 
\hat{{\bf d}}_3\wedge \hat{{\bf d}}_1\right) A_2 - \left(
\hat{{\bf d}}_2 \wedge \hat{{\bf d}}_3 \right) A_1+ \sigma_0
\hat{{\bf d}}_3 \sqrt{16 v_2^2\left( \hat{{\bf d}}_1 \wedge
\hat{{\bf d}}_2\right) ^2-\left( \hat{{\bf d}}_1
A_2+\hat{{\bf d}}_2 A_1 \right) ^2}\right] = 0,
\label{vincd}
\eq
where we have written ${\bf d}_i= d_i \hat{{\bf d}}_i$ ($\hat{{\bf
d}}_i$, $i=1,2,3$ are unit vectors)   
and we have defined $A_1=(v_2^2-v_1^2+4 d_1^2)/d_1$,
$A_2=(v_3^2-v_2^2-4 d_2^2)/d_2$ and $\sigma_0 = sgn({\bf d}_1 \wedge {\bf
d}_2 \cdot {\bf v}_1)$.

The above constraint, together with the requirement that 
all the distances among the helices range between $\rho_0$ and
$\rho_1$, defines the ground state configuration of our protein.

It is straightforward to see that Eq.~(\ref{vincd}) has solutions on
the plane: for instance upon choosing ${\bf d}_3=- {\bf d}_1$, we see
that any configuration satisfying:
\be
\rho_0^2+ \frac{1}{4}\mid v_3^2-v_2^2\mid  \le d_1^2 +d_2^2 \le
\rho_1^2- \frac{1}{4}\mid v_3^2-v_2^2\mid 
\label{solregion}
\ee
 is a solution,
and indeed a bundle-shaped one, because 
of the geometrical conditions we have imposed. 
These solutions share
the same topology and present no barrier in between, since  the
corresponding domain of $(d_1,d_2)$ plane is simply connected
(Eq.~(\ref{solregion})). For these reasons
they can be considered as small deformations of the same "native
state". 
Of course solutions exist which are not planar, but they are
difficult to single out analytically. Hence, we resorted to numerical
calculations to find  some of the energy minima, which, due to 
the  oversimplified, square well
interaction potential Eq.~(\ref{hij}), are highly degenerate.
First, we characterized the conformations of the protein by 
spherical coordinates 
${\bf v}_i \equiv (v_i, \theta_i, \phi_i)$   
calculated with respect to vector  ${\bf v}_1$. Since the lenghts are
known, and we can set $\phi_2=0$, we were left with the five angular
coordinates $\theta_2$, $\theta_3$, $\theta_4$, $\phi_3$, $\phi_4$,
to describe any configuration. We also calculated, for each
configuration, the
``lack of planarity'', as given by the volume $V=\hat{{\bf d}}_1 \wedge
\hat{{\bf d}}_2 \cdot \hat{{\bf d}}_3$, and 
the distance $\beta_{bun}$ 
from the ``most bundle-like'' conformation(the one with
the highest degree of parallelism/antiparallelism  between vectors):
\be
\beta_{bun}= \left[ \frac{1}{6} \sum_{i<j=2}^{4} 
\frac{(\cos(\alpha_{ij})- \cos(\alpha^b_{ij}) 
)^2}{(\cos(\alpha^w_{ij})- \cos(\alpha^b_{ij}) )^2} \right]^\frac{1}{2}\;\;\;.
\label{bundl}
\ee
Here  $\alpha_{ij}$ 
is the angle between vectors ${\bf v}_i$ and ${\bf v}_j$, 
and $\alpha^b_{ij}$, $\alpha^w_{ij}$ are its best and worst values, 
in relation with the bundle configuration.

We performed a random search in configuration space and found nearly $10^4$
ground state configurations; then
we excluded those presenting a
residual steric hyndrance among vectors $\bf v_1$, $\bf v_3$, and 
$\bf v_2$, $\bf v_4$,  since
the condition $\Delta B _{ij} \ge \rho_0$ on the
central points of helices does
not prevent their other points to come too near.

Afterwards we grouped the remaining structures, identifying those
globally differing 
less than $1 \AA$. At the end of this process, we were left with only 4
\% of the original 
configurations, which we sorted according to their $\beta_{bun}$
values.
We found that configurations with the highest degree of
parallelism among the ${\bf v}_i$'s are also those with the most planar set
of vectors ${\bf d_i}$. The best and the worst structure are presented
in Fig.~(\ref{fig:beststr}) and    Fig.~(\ref{fig:worststr}),
respectively: they correspond to the values $\beta_{bun}=0.031$,
$V=0.072$ and   
$\beta_{bun}=0.32$,  $V=0.64$ respectively.  
Notice that, even though our model does not single out a unique native
state for the proposed protein, 
it is nevertheless very encouraging that its results can be easily
pruned, according to some simple considerations of excluded volume, 
leading to essentially  
correct configurations. This,
in spite of the many
simplifying assumptions made.\\

Having investigated the characteristics of the ground state of the
simplified model, we now come to 
the study of its thermodynamic properties, and
write, under certain simplifying assumptions, its partition function.

We consider again the hamiltonian:
\be
H = {\cal V}^0 + {\cal V}^3 + \sum_{i=1}^{N_h} (H_i^0 + H_i^1) \;\;\;.
\label{hsimp}
\ee 
and perform some further modeling on the explicit expression of
$H_i^1$, under the assumption that Eq.(\ref{hi1}) contains more more
details than needed. 
To this purpose we keep 
into account only the
most relevant maxima in $P(l_i,w_i)$, whose positions we specify
by $ (l_{0,\nu}, W_\nu)$.

The general requirements that $H_i^1$ has to fulfil are then the
following:
\begin{itemize}
\item{along the w-axis, it must present minima which are simmetrically disposed
around $\pi$, since ${\bf p}_{\perp }^2$ is invariant under the
exchange $w \rightarrow 2 \pi-w $ (corresponding to the inversion of
handedness of the helix) ;}
\item{in the physically interesting domain, $l_i \in
[q_{1,i},q_{2,i}]$ and $w_i \in [-\pi/3, 5\pi/3]$ (which is the image
of Ramachandran's plane $(\phi,\psi)$ ), $H_i^1$ must amount to a
perturbation of $H_i^0$, hence it must be bounded within a range small
compared to $\gamma_0 c_1 c_3^2 $, namely to the difference between
the maximum and the two (symmetric) minima of $H_i^0$;}
\item{the width, shape and depth of the minima of $H_i^1$ must
resemble those of $P$.}
\end{itemize}

In the following, we shall assume that, near positions $l_{0,\nu}$ along the
sequence, only two minima in $w$ are present, and they are placed at
$\Omega_{1,\nu}= W_{\nu}$ and $\Omega_{2,\nu}= 2 \pi - W_{\nu}$. We
choose for $H_i^1$ the expression:
\be
H_i^1= - n_i \sum_{\nu=1}^{{\cal M}} \sum_{a=1}^2
\gamma_1 \delta_{\nu} G_{i,\nu,a} \;\;\;,
\label{hi1simpl}
\ee
where we have defined:
\be
G_{i,\nu,a} = \exp \{-[\frac{ (l_i - l_{0,\nu} )^2} {\eta_\nu^2} +
\frac{ (w_i - \Omega_{a,\nu})^2 }{\tau_\nu^2}] \}
\label{ginua}
\ee

$H_i^1$ amounts to a collection of $2 {\cal M}$ gaussian wells, whose
overlap will be considered negligible for all practical purposes,
which have a depth $\gamma_1 \delta_{\nu}$ (with $\gamma_1 \ll
\gamma_0 c_1 c_3^2 $), are centered at positions $l_{0,\nu}$ along the
$l$-axis and positions $\Omega_{a,\nu}$ along the w-axis, and have
widths $\eta_\nu / \sqrt{2}$, $\tau_\nu / \sqrt{2}$. We also assume that, 
at the end points of the chain, all the gaussians Eq.(\ref{ginua}) are 
negligible, which will allow us to extend the integrations to the range
$l_i \in [-\infty,\infty]$.

Notice that, while $\gamma_1$ is a tunable parameter, $\delta_{\nu}$
are given (sequence dependent) positive constants, corresponding to
the ratio between the depth of the $\nu$-th minimum and that of
the deepest one. Without loss of generality we can take
$\delta_{\nu} \le 1$.

Expression (\ref{hi1simpl}), (\ref{ginua}) for the hamiltonian relies
on the implicit assumption that the height, position and width of the
maxima of $P(l,w)$ are more important than the details of
its shape, so that a coarse grained description is sufficient. Of
course, the use of gaussians is arbitrary, and is dictaded by the fact
that they allow us to model accurately the shape of $P$ around its
maxima, and decrease rapidly to zero, preventing or reducing spurious
overlaps.

In order to perform the calculations, it is useful to write the
constraint ${\cal V}_3$ appearing in Eq.~(\ref{hsimp}) as:
\be
{\cal V}^3 = \lambda_3 \sum_{i=1}^{N_h} (l_i - L_i)^2\;\;\;,
\label{vinc3q}
\ee
where we have introduced, remembering the definition of $l_i$:
\be
L_i = \frac{1}{2} \left(s_i + s_{i+1} +1 \right) = \frac{n_i+1}{2} +
\sum_{k=1}^{i-1} n_k\;\;\;. 
\ee

Our goal is to evaluate the partition function,
so we write
\be
e^{-\beta H} = e^{\beta ({\cal V}^0 + {\cal V}^3)}
\prod_{i=1}^{N_h} e^{- \beta (H_i^0 + H_i^1 )}\;\;\;,
\ee
and introduce two approximations, in order to make an analytic integration
possible.  
Namely  we replace:
\bq
\lefteqn{\exp \left\{-\beta (n_i-1) \gamma_0 c_1 [(w_i -c_2)^2 -c_3]^2 \right\}
\cong} \;\;\;\; \nonumber \\ 
& & \cong f_i^0 \equiv 
\exp \left\{-\beta (n_i-1) \frac{(w_i- \mu_1)^2}{\sigma^2}\right\} +
\exp \left\{-\beta (n_i-1) \frac{(w_i- \mu_2)^2}{\sigma^2} \right\}\;\;\;,  
\label{approxh0}
\eq
where
\be
\mu_1 = c_2 - \sqrt{c_3} \;\;\;,\;\;\;\mu_2 = c_2 +
\sqrt{c_3} \;\;\;,\;\;\;
\sigma^2 = \frac{1}{4 \gamma_0 c_1 c_3}\;\;\;. 
\ee
We also write:
\be
e^{-\beta H_i^1} \cong f_i^1 \equiv 1 + \sum_{\nu=1}^{{\cal M}}
\sum_{a=1}^{2}
\exp \{- \beta e_{\nu}(n_i)
[\frac{ (l_i - l_{0,\nu} )^2} {\eta_{\nu}^2} + 
\frac{ (w_i - \Omega_{a,\nu})^2 }{\tau_{\nu}^2}] \}\;\;\;,
\label{approxh1}
\ee
where
\be
e_{\nu}(n_i) = n_i \gamma_1 \delta_{\nu}
\frac{e^{\beta n_i \gamma_1 \delta_{\nu} } }
{e^{\beta n_i \gamma_1 \delta_{\nu}} -1}\;\;\;.
\label{enu}
\ee
Notice that in this case both $e^{-\beta H_i^1}$ and $f_i^1$ tend to one
as we move far from the maxima: nevertheless, this fact does not
bring any further difficulty  in the integrations involved in the
partition function, since the dependence of $H$ on $w_i$ is dominated by
$H_i^0$, and that on $l_i$
(which anyway ranges between the two finite values $q_{1,1}$ and
$q_{2,N_h}$, see Eq.(\ref{dyva}))
by ${\cal V}^3$.
With the above approximations, which are thoroughly discussed  in
Appendix~\ref{app3},
we are finally able to calculate the partition function. First of all we
integrate on $w_i$ and $z_i$, performing the change of variable $\zeta_i
= z_i-c_6 + c_7 (w_i-c_8)^2$, which does not affect the jacobian: 
\[
Z(\{n\},\{l\}) = e^{-\beta ({\cal V}^0 +{\cal V}^3)} \prod_{i=1}^{N_h}
\int_{-\infty}^{\infty} dw_i d\zeta_i f_i^0 f_i^1
\exp[-\beta \gamma_0 c_5 (n_i -1) \zeta_i^2]\;\;\;.
\]

We insert Eqs.(\ref{approxh0}, \ref{approxh1}) in the above expression
and perform the integrations on $\zeta_i$, $w_i$
and then the summations on $l_i$, which we extend to the range
$[-\infty,\infty]$. Eventually, after some lenghty calculations,
and resorting to the definition of elliptic theta 
functions\cite{hans}, 
we can write:

\bq
Z(\{n\}) = e^{-\beta {\cal V}^0 }  \prod_{i=1}^{N_h} \!\!\!&{}&\!\!\!
 \left\{    \frac{2 \sigma \pi}{\beta (n_i -1) \sqrt{\gamma_0 c_5}}
        \theta_3(0,e^{-\frac{1}{4} \beta \lambda_3} )+ \right. 
 \left. \sum_{\nu=1}^{{\cal M}} \sum_{a,b=1}^{2}
          \left[ d_{\nu}(n_i)
    e^{\left(A_{a,b,\nu}(n_i) - C_\nu(\{n_k\}, \lambda_3)\right)}
 \times \right. \right.
\nonumber\\
                &\times& \left. \left.\frac{\sqrt{\pi}}{x_\nu(n_i,\lambda_3)}
                   \theta_3 \left(\pi \frac{y_\nu(\{n_k\},\lambda_3)}
                               {x_\nu(n_i,\lambda_3)},
                                e^{-\pi^2 x_\nu^{-2}(n_i,\lambda_3)} \right)
\right] \right\} \;,
\label{znteta}
\eq
where $\theta_3(z,q)$ is the elliptic $\theta_3$ function of argument $z$ and
nome $q$, and we have defined:
\bqn
d_{\nu}(n_i) &=& \frac{\pi}{\beta} (e^{\beta n_i \gamma_1 \delta_{\nu}} -1)
             \left[(n_i -1) \gamma_0 c_5
                 (\frac{n_i -1}{\sigma^2}+
             \frac{e_{\nu}(n_i) }{\tau_{\nu}^2})
                  \right]^{-\frac{1}{2}} \;\;\;,\\
A_{a,b,\nu}(n_i) &=& \frac{\beta}{\sigma^2 \tau_{\nu}^2} \left\{
   \frac{\left[(n_i -1) \mu_a \tau_{\nu}^2 +  
                e_{\nu}(n_i) \Omega_{b,\nu} \sigma^2  \right]^2}
        {(n_i -1)  \tau_{\nu}^2 +  
         e_{\nu}(n_i) \sigma^2 } -
   \left[(n_i -1) \frac{\mu_a^2} {\sigma^2} +
         \frac{ e_{\nu}(n_i)}{\tau_{\nu}^2}
         \Omega_{b,\nu}^2 
  \right] \right\} \\
x_\nu(n_i,\lambda_3) &=& \frac{1}{2 \eta_{\nu}} \sqrt{\beta (e_\nu(n_i) +
\lambda_3 \eta_{\nu}^2)}
               \;\;\;,  \\
y_\nu(\{n_k\},\lambda_3) &=&  
\sqrt{\frac {\beta \eta_{\nu}^2} {e_\nu(n_i) +
\lambda_3 \eta_{\nu}^2}}
         (l_{0,\nu} \frac{e_\nu(n_i)}{\eta_{\nu}^2} + \lambda_3 L_i(\{n_k\})) 
\;\;\;, \\
C_\nu(\{n_k\},\lambda_3) &=&
        \beta \frac{e_\nu(n_i) \lambda_3}{e_\nu(n_i) + \lambda_3 \eta_{\nu}^2}
           (l_{0,\nu} - L_i(\{n_k\}))^2 \;\;\;,
\eqn
At this point we should perform the sums over $n_i$. They are clearly
unfeasible in  analytical way, yet, they are very simple to perform
numerically. To this end it may be useful to implement directly the
constraint ${\cal V}_0$ by setting
\[n_{N_h} = N - \sum_{k=1}^{N_h -1} n_k \;\;\;,\]
and paying the necessary attention to discard the collection of
$n_k$'s leading    to  negative values for $n_{N_h}$.

In this way, for each given sequence one can obtain an expression for
the  partition function which depends only on $\beta$, $\lambda_3$, and
the ratio $\gamma_1/\gamma_0$.
The Lagrange multiplier must be evaluated by minimizing the free energy.
This involves the condition
\[ \frac{\partial Z}{\partial \lambda_3} =0 \;\;\;,\]
that should be solved to give $\lambda_3 = \lambda_3(\beta,
\gamma_1/\gamma_0)$.

We shall then be able to study the thermodynamic behaviour of the system
at different temperatures and values of $\gamma_1/\gamma_0$. We
leave this detailed analysis to future studies, since our goal here was
only two show that, despite the complexity of the model, interesting
calculations, providing  new insight in  the folding process, can be
performed without resorting to heavy numerical work.

\section{Comments and conclusions}
\label{sec-conc}
This paper is mainly devoted to the presentation and analysis of a new
model hamiltonian to be used in thermodynamical and dynamical studies of the
folding process. The various contributions to the hamiltonian
(\ref{protham}), and their relation to phenomenology,
have been thoroughly discussed, together with the approximations introduced.

Then we concentrated on the local part of the hamiltonian, 
and applied it to the study of
a small (73-residues long) synthetic protein \cite{rade92}, 
designed to produce a four-helix bundle. 

Even if this part of the model does not
contain information on the spatial structure, it takes into 
account the sequence, so we expect that it can 
provide relevant information 
on  the secondary structure composition of the native state.
To prove that this is indeed the case, 
we studied 
the local hamiltonian minima for the protein considered,
 and then switched on
the non-local interactions as a small perturbation, thus superimposing
the tertiary structure to the existing secondary one.  

We found that  the four-helices ground-state
is correctly made up of $\alpha$-helices, placed in the same 
positions along the
chain as they  are found in the experimental native state.
Encouraged by this  result, we studied the best spatial configuration 
that the four helices we obtained would assume, due to the non-local
mutual interactions. Again we obtained a positive result:
bundle-like configurations are indeed the  ground-state of the model,
even if the latter appears to be degenerate, due to the
small number of helices involved  and to the
highly simplified expression of the interaction hamiltonian $H_{ij}$.
It is very likely that the degeneracy would be partially or completely
removed by the introduction of an interaction potential
depending not only on the
distance between helices, but also on the mutual orientation of both the
hydrophobic moments and  of the helices themselves.

Finally, resorting to some simplifying assumptions, we showed
that  the partition function of the simplified, local model can 
be evaluated almost
completely in  analytic way. 

A  detailed study of natural proteins with the above approach is left to
future work.
As already mentioned, we also leave to future efforts the 
refinements regarding, for istance, the explicit expressions for $P$,
appearing in Eq.~(\ref{hi1}), and for the charges (\ref{qbar}). In this 
paper they have been fixed in an arbitrary, though reasonable, way 
just to perform some tests on the 
model. The same holds true for assumption (\ref{hi1simpl}), whose validity
could depend on the sequence considered and requires further
analysis. 
    
Coming to the partition function for the complete model, 
a reasonable goal is to perform exact,
analytic integration on some of the variables (for instance, $w_i$,
$z_i$, $l_i$), thus providing an effective interaction potential among
the others; then, one could resort to numerical simulations.
The latter could be approached, for instance, putting the ${\bf B}_i$ on a
lattice: in this way one could have a true mapping of real proteins on
lattice models, and the approximations induced by the lattice could be
better controlled in their relationship to protein geometry.

The fact that the known native state of a real protein can be mapped
onto a model configuration entails  also that the study of the
inverse folding problem can greatly benefit from this new approach.

Finally, it is easy to provide a coarse-grained dynamics for the
protein  through the
variables used in the model, and our future efforts will be dedicated
also to this line of research.

\acknowledgements
We thank Mario Rasetti, Vittorio Penna and Riccardo Zecchina for 
helpful discussions, and Chao Tang for kindly providing us with the data 
about $q_i$ values.

\appendix

\section{} 
\label{app1}
Curvature and torsion of the curve $r(s)$ are defined:
\begin{equation}
\kappa = \frac{\mid {\bf \dot{r} \wedge \ddot{r}} \mid}{\mid {\bf      
\dot{r}} \mid^3}\;\;\;,\;\;\;
\tau = \frac{ {\bf \dot{r} \wedge \ddot{r} \cdot
\overdots{{\bf r}} } }
{\left| {\bf \dot{r} \wedge \ddot{r}} \right|^2} \;\;\;.
\nonumber
\end{equation}

Performing the calculations for the curve in Eq.(\ref{chain}), and keeping only
the leading terms in the limit $\lambda \rightarrow 0$, we find:
\begin{eqnarray*}
\mid {\bf \dot{r}}\;\; &\mid \stackrel{\rm \lambda \rightarrow 0}{=}& 
\left[ \sum_{i=1}^{N_h} 
\left[ \left( b_i \dot{{\bf h}}_i + b_{i+1} \dot{{\bf h}}_{i+1}
\right)^2 - 
b_{i+1}^2 \dot{{\bf h}}_{i+1}^2 \right] \right]^{\frac{1}{2}} \;\;\;, \\
\left| {\bf \dot{r} \wedge \ddot{r}} \right|
&\stackrel{\rm \lambda \rightarrow 0}{=}&
\left[ \sum_{i=1}^{N_h}
\left( \left[\dot{g_i} (b_i + b_{i+1})(\dot{{\bf h}}_i \wedge \dot{{\bf
h}}_{i+1}) +  (b_i \dot{{\bf h}}_i + b_{i+1} \dot{{\bf h}}_{i+1}) 
\wedge (b_i \ddot{{\bf h}}_i + b_{i+1} \ddot{{\bf h}}_{i+1}) 
\right]^2 -  \right. \right. \\
&& \left. \left. b_{i+1}^4 (\dot{{\bf h}}_{i+1}
\wedge \ddot{{\bf h}}_{i+1})^2
\right) \right]^{\frac{1}{2}} \;\;\;,\\ 
{\bf \dot{r} \wedge \ddot{r} \cdot 
\overdots{{\bf r}} } 
&\stackrel{\rm \lambda \rightarrow 0}{=}& 
\sum_{i=1}^{N_h} \left[ b_i^3 \dot{{\bf h}}_i \wedge \ddot{{\bf h}}_i 
\cdot 
\overdots{{\bf h}}_i + 
(b_i - b_{i+1}) \left[6 \dot{g}_i^2 
                              (\dot{{\bf h}}_i \wedge \dot{{\bf h}}_{i+1}) 
(- \ddot{{\bf h}}_i + \ddot{{\bf h}}_{i+1}) + \right. \right.\\
&& \left.\left. 2 \dot{g}_i (b_i 
\overdots{{\bf h}}_i + 
      b_{i+1} 
\overdots{{\bf h}}_{i+1}) \wedge \dot{{\bf h}}_i
\cdot \dot{{\bf h}}_{i+1}+ 
 3 \dot{g}_i (b_i \dot{{\bf h}}_i + b_{i+1} \dot{{\bf h}}_{i+1})
\wedge \ddot{{\bf h}}_i \cdot \ddot{{\bf h}}_{i+1} \right] + \right.\\
&&\left. \frac{1}{2} \sum_{\mu, \nu, \rho=1}^3 \varepsilon_{\mu \nu \rho} 
b_i b_{i+1} (b_i {\bf h}_i^{(\mu)} + b_{i+1} {\bf h}_{i+1}^{(\mu)})
\wedge {\bf h}_i^{(\nu)} \cdot {\bf h}_{i+1}^{(\rho)} \right]\;\;\;.  
\end{eqnarray*}
In the above equations, dots as well as greek apices indicate 
derivative  with respect to $s$, and we have introduced ${\bf h}_0= 
{\bf h}_{N_h+1} = {\bf 0} $.

Now we observe that the above quantities, in the limit
considered, are made up by terms proportional to powers of $b_i$, that
are different from zero in the region $s_{i-1} < s < s_i$, and terms
that live in the interfaces $s=s_i$ between helices, namely those
containing the product $b_i b_{i+1}$, or the delta function $\dot{g}_i$.
These terms are mutually ``orthogonal'', in the sense their product
is zero. This fact allows strong simplifications,
because one can perform algebraic manipulations at fixed s (this is
allowed, since we are not differentiating), and consider only those
terms which are different from zero for that particular value of s.
For instance, the curvature becomes:
\begin{equation}
\kappa = \sum_{i=1}^{N_h} \left( \kappa_i \vartheta(s-s_{i-1}) \vartheta(s_i-s)
+ 2 q_i(s) \frac{\mid 4 \delta(s-s_i) (\dot{{\bf h}}_i \wedge 
\dot{{\bf h}}_{i+1}) +
(\dot{{\bf h}}_i + \dot{{\bf h}}_{i+1}) \wedge 
(\ddot{{\bf h}}_i + \ddot{{\bf h}}_{i+1}) \mid}
{\mid \dot{{\bf h}}_i + \dot{{\bf h}}_{i+1} \mid^3} \right),
\nonumber
\end{equation}
where $\vartheta$ is Heaviside's function, $\kappa_i$ is the curvature
of  the i-th helix:
\[ \kappa_i = \frac{\mid \dot{{\bf h}}_i \wedge \ddot{{\bf h}}_i \mid}
{\mid \dot{{\bf h}}_i \mid^3 } \;\;\;,\]
and the delta-like functions \[q_i(s) = \left\{ \begin{array}{c}
                         1 \;,\; {\textstyle if}\;\; s = s_i\\ 
                         0 \;,\;{\textstyle otherwise}
                       \end{array}  \right. \] 
have been introduced to single out those terms that live at the
interfaces $s_i$.
Obviously, the term containing $\delta(s-s_i)$ is the only one 
to take into account, which leads to the result in Eq.(\ref{chaink}).

\section{}
\label{app2}
In order to study the relationship between geometrical quantities
and ($\phi,\psi$) angles, we observe that a protein may be built up by
performing a sequence of rotations and translations of the peptide plane
$C_i^{\alpha} C' N C_{i+1}^{\alpha}$.  

If  ${\bf L}$ is the segment joining to successive $C^{\alpha}$, and
if we  label
$j$ the peptide plane between $C_j^{\alpha}$ and $C_{j+1}^{\alpha}$, we have
that the position of $C_N^{\alpha}$ is given, in the reference frame of peptide
plane number 0, by:
\bq
{\bf r}_0(C^{\alpha}_N) &=& \left[ \prod_{i=1}^{N-1} ({\bf L} + R(\phi_i,
\psi_i))   \right]{\bf L} \nonumber\\
&=&  {\bf L} + R(\phi_1,\psi_1){\bf L} + \; \cdots \;+ 
             R(\phi_1,\psi_1) \cdots R(\phi_{N-1},\psi_{N-1}){\bf L} \;\;\;.
\label{rototr}
\eq 
Following Ramachandran \cite{rasa68}, we associate to each peptide plane $j$ a
reference frame $({\bf x}_j,{\bf y}_j,{\bf k}_j)$, such that the origin sits on 
$C^{\alpha}_j$, $ {\bf y}_j =\hat{{\bf L}}_j$ and ${\bf x}_j$ lies on the 
plane,
with ${\bf x}_j \cdot \vec{C'O} > 0$.

Now we consider two successive planes, labelled 0 and 1, and take $\hat{\phi},
\hat{\psi}$ to be the unit vectors, expressed in reference frame 0, of the
rotation axes $N C^{\alpha}_1$ and $C^{\alpha}_1 C'$, when $\phi = \psi = 0$   
(according to the standard conventions, this corresponds to having
$C' N$, $N C^{\alpha}_1$, $C^{\alpha}_1 C'$ in "{\it cis}"
configuration,  lying on the
same plane with $\vec{C'N} \cdot  \vec{C^{\alpha}_1 C'} < 0$).

Given frame 0, the sequence of rotations to perform in order to obtain frame 1
is the following: first one rotates by an angle $\pi$ about the ${\bf y} $
axis, then of an angle $\theta= -\pi + 
C^{\alpha}_0\hat{C^{\alpha}_1}C^{\alpha}_2 = -(\pi - \mid \alpha \mid + \mid 
\zeta \mid +
\mid \beta \mid)$ about the ${\bf z}$ axis, to recover the
standard $\phi = \psi = 0$ configuration. Here $\alpha = N\hat{C}^{\alpha} C'$, 
$\beta = C'\hat{C}^{\alpha}_1 C^{\alpha}_2$, and 
$\zeta= C^{\alpha}_0\hat{C}^{\alpha}_1 N$.
Then we can perform the $\psi$-rotation, and successively the
$\phi$-one, obtaining:
\be
{\bf X}^1_a = {\bf X}^0_b [R(\mbox{\boldmath $\phi$})
R(\mbox{\boldmath $\psi$}) R(\theta {\bf z}) R(\pi {\bf y})  ]_{ba}\;\;\;,
\label{rot}
\ee
where all the rotation axes are expressed in reference frame 0, 
$({\bf X}^j_1 , {\bf X}^j_2 , {\bf X}^j_3) = ({\bf x}_j,{\bf y}_j,{\bf
z}_j)$, 
and
\be
[R(\mbox{\boldmath $\eta$})]_{pq} = \cos(\eta) \delta_{pq} + (1-\cos(\eta))
\eta^p \eta^q - \sin(\eta) \eta^r \varepsilon^{rpq}
\label{eachrot}
\ee
(sum over repeated indices is understood; $\eta^p$ are the cosine directors of
rotation vector \mbox{\boldmath $\eta$}).

The above product of
rotations is equivalent to a single rotation\cite{vamoa} 
$R(\mbox{\boldmath$\omega(\phi,\psi)$})$ 
with the argument \mbox{\boldmath$\omega(\phi,\psi)$} satisfying:
\bq
c \equiv c(\omega) &=& \hat{\mbox{\boldmath $\kappa$}} \left( {\bf s}(\psi)
c(\phi) + {\bf s}(\phi)c(\psi) \right) \;\;\;,
\label{cosom}
\\
{\bf s} \equiv {\bf s}(\omega) &=& 
\hat{\mbox{\boldmath $\kappa$}} \left[ {\bf s}(\psi) {\bf s}(\phi) - 
c(\phi) c(\psi) \right] - \hat{\mbox{\boldmath $\kappa$}} \wedge 
({\bf s}(\psi) \wedge {\bf s}(\phi))+ \hat{\mbox{\boldmath $\kappa$}} \wedge
\left( {\bf s}(\psi) c(\phi) + {\bf s}(\phi)c(\psi) \right)\;\;\;, 
\label{sinom}
\eq
where, for any argument $\eta$, $c(\eta) = \cos(\eta/2)$, ${\bf s}(\eta) = 
\hat{\mbox{\boldmath $\eta$}} \sin(\eta/2)$; while

\bqn
\hat{\mbox{\boldmath $\kappa$}} &=& (-\sin(\frac{\theta}{2}), 
\cos(\frac{\theta}{2}),0) \;\;\;,\nonumber \\
\hat{\mbox{\boldmath $\phi$}} &=& (\sin(\mid\zeta\mid), \cos(\mid\zeta\mid),0)
\;\;\;,\nonumber \\
\hat{\mbox{\boldmath $\psi$}}  &=& (\sin(\mid\alpha\mid - \mid\zeta\mid), 
-\cos(\mid\alpha\mid - \mid\zeta\mid),0) \;\;\;,
\eqn

The above equations specify completely the rotation matrices appearing in 
Eq.(\ref{rot}).

Now we have to relate the parameters identifying a helix, to the
repeating value of the angle of rotation 
\mbox{\boldmath $\omega(\phi,\psi)$}.
Let us take three successive  $C^{\alpha}-C^{\alpha}$ segments, namely
${\bf L}$, ${\bf L}' $, $ {\bf L}'' $ , and put ourselves in 
reference frame of
segment ${\bf L}' $. We have ${\bf L} = R^{-1}(\phi,\psi) {\bf L}' $ and 
${\bf L}'' = R(\phi,\psi) {\bf L}'$ , whence
\bqn
{\bf L} &=& L \{2(s_1 s_2 +c s_3), c^2-s^2 + 2 s_2^2, 2 (s_2 s_3 -c
s_1)\} \;\;\;,\\
{\bf L}' &=& L \{0,1,0\} \;\;\;,\\
{\bf L}'' &=& L \{2(s_1 s_2 -c s_3), c^2-s^2 + 2 s_2^2, 2 (s_2 s_3 +c s_1)\}
\;\;\;,\eqn
where $c=c(\omega)$, $s=\sin(\omega/2)$, $s_i= \omega_i
\sin(\omega/2)$ ($s_i$ are the components of the vector ${\bf s}(\omega)$).

Now we ask that the helix axis ${\bf e}_3$ satisfies the following 
requirements:
\be
{\bf L \cdot e}_3 = {\bf L' \cdot e}_3 \;\;\;\;,\;\;\;\;
{\bf L'' \cdot e}_3 = {\bf L' \cdot e}_3  
\label{cond1}
\ee
The axis orientation is fixed by:
\be
{\bf L \cdot e}_3 > 0 \;\;\;,
\label{cond2}
\ee
and, to distinguish between left- and right-handed helices, we
assume that:
\be
sgn({\bf L \wedge L' \cdot e}_3) = \cases{+1, &if right-handed \cr
-1, &if left-handed.}
\label{cond3}
\ee

We need also to relate the lenght of the helix arc, 
corresponding to one peptide segment $L$ of the chain, to the angle of
rotation at each site.
From the definition of $u$ (Eq. (\ref{u})), we have that $\Delta s =1$
corresponds to an arc of lenght $L$; hence $u$ is the rotation about the
helix axis corresponding to a displacement of one peptide unit along
the chain. This angle must be equal to the projection of $\omega$ on
the plane perpendicular to ${\bf e}_3$, namely 
\be
\cos(u)= \frac{{\bf e}_3 \wedge ({\bf L} \wedge {\bf e}_3)}
{\mid {\bf L} \wedge {\bf e}_3 \mid} \cdot  
\frac{{\bf e}_3 \wedge ({\bf L}' \wedge {\bf e}_3)}
{\mid {\bf L}' \wedge {\bf e}_3 \mid} \;\;\;.
\label{cosu}
\ee

Equations (\ref{cond1})-(\ref{cond3}) yield:
\be
{\bf e}_3 = \sigma \frac{{\bf s}}{\mid s \mid}\;\;\;,
\ee 
where $\sigma = \pm 1$. Observing that $\omega \in [0, 2\pi]$ 
($\omega \in [-\pi, \pi]$ would not be correct, as $c(\omega)$ in Eq.
(\ref{cosom}) can also be negative), we always have $s>0$, hence 
\be
{\bf e}_3 = \sigma \hat{\mbox{\boldmath $\omega$}}\;\;\;. 
\ee 

From Eq.(\ref{cosu}) we finally get:
\[
\cos(u) = \cos(\omega)\;\;\;.
\] 

To establish the sign $\sigma$ and the explicit expression of $u$, we
use Eqs.(\ref{cond2}),(\ref{cond3}), and find that the product $s_2 c$
determines the handedness of the helix (right-handed if positive),
while $\sigma$ coincides with $sgn(s_2)$ 
whenever $s_2$ is different from zero. The analysis of the
case $s_2 = 0$ lead us to the following general definition:
\be
u= \sigma(s_2) \left( \omega - 2 \pi \left[\!\left[\frac{\omega}{\pi}\right]\!
\right]  \right)\;\;\;,
\ee 
where $[\![ x ]\!]$ denotes the maximum integer $ \le x$, and $\sigma(x) =
1$ if $x > 0$, $\sigma(x)=-1$ otherwise.
With the above position we have $u \in [-\pi, \pi]$, with positive
values  corresponding to
right-handedness. Recalling definition (\ref{w}) we can also write: 
\be
w= \pi + \sigma(s_2) \left( \omega - \pi \right)\;\;\;.
\ee 

Now we can derive $a$ and $h$, from Eq.(\ref{ahu}) and the condition
\be 
a h u = {\bf L \cdot e}_3 = L  \left| \frac{s_2}{s} \right|\;\;\;,
\ee
and hence, using once more Eq.(\ref{ahu}), curvature and torsion. The former
can be expressed as
\[ \kappa = \frac{\mid u \mid}{L} \sqrt{1-\frac{\tau^2 L^2}{u^2}}
\;\;\;,\]
while torsion is given by
\be
\tau = \left| \frac{s_2}{s} \right| \frac{u}{L}\;\;\;.
\label{tors}
\ee

Now we recall that $\omega$ depends on $(\phi, \psi)$ through
Eqs.(\ref{cosom}) and (\ref{sinom}), where $\cos(\frac{\omega}{2})$ and
$\hat{\mbox{\boldmath $\omega$}} \sin(\frac{\omega}{2})$ appear. 
From Eq.(\ref{tors}) and the relation $\cos\frac{u}{2}= \mid
\cos\frac{\omega}{2} \mid$, using the explicit expressions of the
vectors appearing in Eq.(\ref{sinom}), we find two formulas relating $u$ and
$\tau$ to the dihedral angles: 
\[
\cos\frac{u}{2} =  \left| c(\omega)  \right| \;\;\;,\;\;\; 
z  = \frac{\left| s_2 \right |}{\sqrt{1 - c^2(\omega)}} \;\;\;,  
\]
where the quantities at the right hand side of the above equations can be
explicitly written as
\bq
c(\omega) &=& 
\sin \frac{\mid \alpha \mid}{2} 
\cos(\frac{\mid \zeta \mid - \mid \beta \mid}{2}) 
\sin(\frac{\phi + \psi}{2}) 
+ \cos\frac{\mid \alpha \mid}{2}
\sin(\frac{\mid \zeta \mid - \mid \beta \mid}{2}) 
\sin(\frac{\psi -  \phi}{2}), \nonumber \\
s_2 &=&  
\sin \frac{\mid \alpha \mid}{2} 
\cos(\frac{\mid \zeta \mid + \mid \beta \mid}{2}) 
\cos (\frac{\psi +\phi}{2}) 
+ \cos\frac{\mid \alpha \mid}{2}
\sin(\frac{\mid \zeta \mid + \mid \beta \mid}{2}) 
\cos(\frac{\psi -\phi}{2}) . \nonumber 
\eq
In analogous  way we find for  $w$, $z$:
\bqn
\cos\frac{w}{2} &=&  \sigma(s_2) c(\omega)  \;\;\;, \;\;\; \\
z  &=& \frac{\left| s_2 \right |}{\sqrt{1 - c^2(\omega)}} \;\;\;.  
\eqn
which is identical to Eq.(\ref{inv}), where
an obvious definitions of coefficients $a$,$b$,$c$,$d$ has been performed.

\section{}
\label{app3}
In this appendix we discuss the validity of the two approximations
Eqs.(\ref{approxh0},\ref{approxh1})  we
introduced in section~\ref{sec-simpmod}.
The first approximation is the most relevant one, since it involves the
leading hamiltonian $H_i^0$. The parameters $\mu_1$, $\mu_2$, $\sigma$
have been chosen so that the exact and approximate
functions have maxima at the same positions, and the leading order
in the series expansion at those points are the same, provided that
the overlap between the two gaussians is vanishing:
\[ \exp[-\beta (n_i-1)\frac{(\mu_2-\mu_1)^2}{\sigma^2}] \approx 0 \;\;\;.\]  

We check that the approximation is globally good by a comparison of the
two integrals:
\bq
{\cal I}_1 &=& \int_{-\infty}^{\infty} dw_i e^{-\beta (n_i-1)
\gamma_0 c_1 [(w_i -c_2)^2 -c_3]^2} \;\;\;, \label{intp4} \\
{\cal I}_2 &=& \int_{-\infty}^{\infty} dw_i
\left\{ e^{-\beta (n_i-1) \frac{(w_i- \mu_1)^2}{\sigma^2}} +
e^{-\beta (n_i-1) \frac{(w_i- \mu_2)^2}{\sigma^2}} \right\}
 \nonumber \\
&=&  \sqrt{\frac{\pi c_3}{2}} \frac{1}{\sqrt{y}}\;\;\;,
\label{intgaussh0}
\eq
where $y= \beta (n_i-1) \gamma_0 c_1 c_3^2/2$.

The first one can be evaluated explicitly:
\be
{\cal I}_1 = \frac{\pi}{2} \sqrt{c_3} e^{- y }
\left[ I_{-\frac{1}{4}}(y) + I_{\frac{1}{4}}(y)
\right]\;\;\;,
\label{intp4sol}
\ee
where $I_{\pm \frac{1}{4}}(y)$ are modified Bessel function of the
first kind.

A numerical evaluation of
\be
\epsilon = \left| 1- \frac{{\cal I}_2}{{\cal I}_1} \right|
\label{relerr}
\ee
reveals that, for $y \gsim 0.24$ 
it always holds $\epsilon \lsim 0.12$, with $\epsilon$ decreasing to zero as
$y$ increases.
 
For the above reasons we consider expression Eq.(\ref{approxh0}) as a good
approximation, with a word of caution: the value of its right and left
member differ significantly at $w_i = c_2 = (\mu_1 +\mu_2)/2$, the
maximum of the quartic potential appearing in $H_i^0$.
We have in fact that the former   
goes as $\exp[-2 y]$, while the latter gives $2 \exp[-8 y]$.
This difference may however be considered negligible when both of the above
quantities are very
small, namely, for long enough helices and/or low enough temperature
$1/\beta$.  
   
Coming to the second approximation, we see that also in this case Taylor
expansions near the maxima $(l_{0,\nu},\Omega_{a,\nu})$ coincide.
To check the global behaviour, we proceed as before, evaluating: 
\be
{\cal I}_3 = \int_{-\infty}^{\infty} dw_i dl_i (e^{-\beta H_i^1}-1)\;\;\;, 
\label{inth1 }
\ee
moving first to polar coordinates:
\[\frac{l_i -l_{0,\nu}}{\eta_\nu} = r \cos(\theta)\;\;\;\;,\;\;\;\;
  \frac{w_i -\Omega_{a,\nu}}{\tau_\nu} = r \sin(\theta) \;\;\;,\]
then introducing  $z=\exp(-r^2)$,  and finally resorting to formula  
Eq. 5.1.40 in the Abramowitz-Stegun handbook\cite{abst}, whereby: 
\be
{\cal I}_3 = \eta_\nu \tau_\nu \pi [Ei(b)- \ln(b) - \gamma]\;\;\;,
 \label{inth1sol }
\ee
where $\gamma$ is the Euler constant, $Ei$ indicates the exponential
integral function, and $b= \beta n_i \gamma_1
\delta_{\nu}$. 

This is to be compared to:
\be
{\cal I}_4 =  \int_{-\infty}^{\infty} dw_i dl_i (f_i^1 -1) 
           =  4\frac{ \eta_\nu \tau_\nu \pi}{b} \sinh^2(\frac{b}{2})\;\;\;.
\label{int4}
\ee
Again, an estimate of $\epsilon' = \left| 1- {\cal I}_4/{\cal I}_3
\right| $ reveals that $\epsilon' \rightarrow 0$ both for $b \rightarrow
0$ and for $b \rightarrow \infty$, with a maximum of $\epsilon'=0.27$ at
$b=2.97$.


\begin{figure}
\caption{Contour lines of $\frac{H^0_i}{(n_i-1) \gamma0}$, showing the two
minima in ($w,z$) plane; the variable $w$, on the horizontal axis ranges from
$\pi/3$ to $5 \pi/3$, while $z$, on the vertical axis, goes from 0 to 1.
Lines are drawn at intervals of 1, in arbitrary units, in the range
[0,5] (deeper regions are darker). The saddle point corresponds to
an height of 2.7.
The figure is obtained with the values $c_1= 2.73$, $c_2=2.77$, 
$c_3= 0.99$, $c_4= 0$, $c_5=150$, $c_6=0.86$, $c_7= 0.16$, $c_8= 3.5$.}
\label{figv4wz}
\end{figure}

\begin{figure}
\caption{The same as  
in Fig.(1),
but with $w$, $z$
expressed as functions of $\phi$ (horizontal axis) and $\psi$ (vertical axis)  
through Eq.(14).
Both $\phi$ and $\psi$ range from
$-\pi$ to $\pi$. The values of the parameters are the same as in Fig.(1).
Notice the essentially correct position and shape of the $\alpha$ and
$\beta$ minima; the unphysical third minimum is an effect of the
approximate symmetry of 
Eq.(14)
 under the transformation 
$\phi \leftrightarrow \psi$.}

\label{figv4fipsi}
\end{figure}

\begin{figure}
\caption{Plot of $P(l,w_{0 \alpha})$ (continuous line), $P(l,w_{0
\beta})$ (dotted line) and $0.1 \overline{q}(l)$ (dashed line, below)
for the considered protein, calculate for a 7-residues-long helix centered at 
position l (on the horizontal axis);
$\overline{q}$ and $P$ have 
been defined in Eqs. (12, 26). 
Notice the big 
maxima of the dipole moment  characterising the $\alpha$ helices.}
\label{figp2albe}
\end{figure}
\begin{figure}
\caption{Ground-state configuration most similar to a 
bundle, among those obtained by numerical 
calculations: $\beta_{bun}=0.031$, $V=0.072$ (see text).}
\label{fig:beststr}
\end{figure}
\begin{figure}
\caption{Ground-state configuration with the worst value of $\beta_{bun}$ 
and $V$, among those obtained by numerical calculations: 
$\beta_{bun}=0.32$, $V=0.64$.}
\label{fig:worststr}
\end{figure}

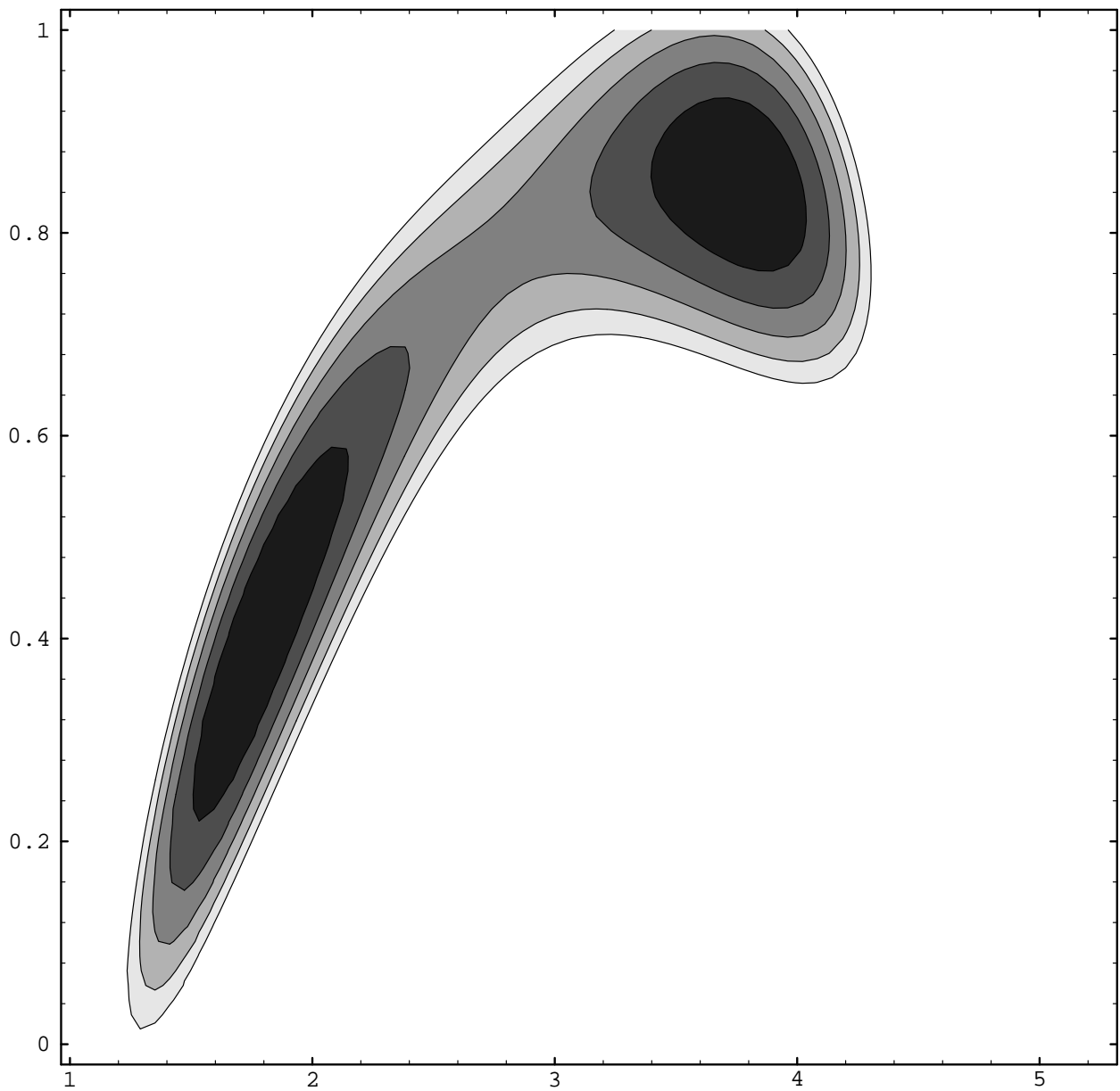
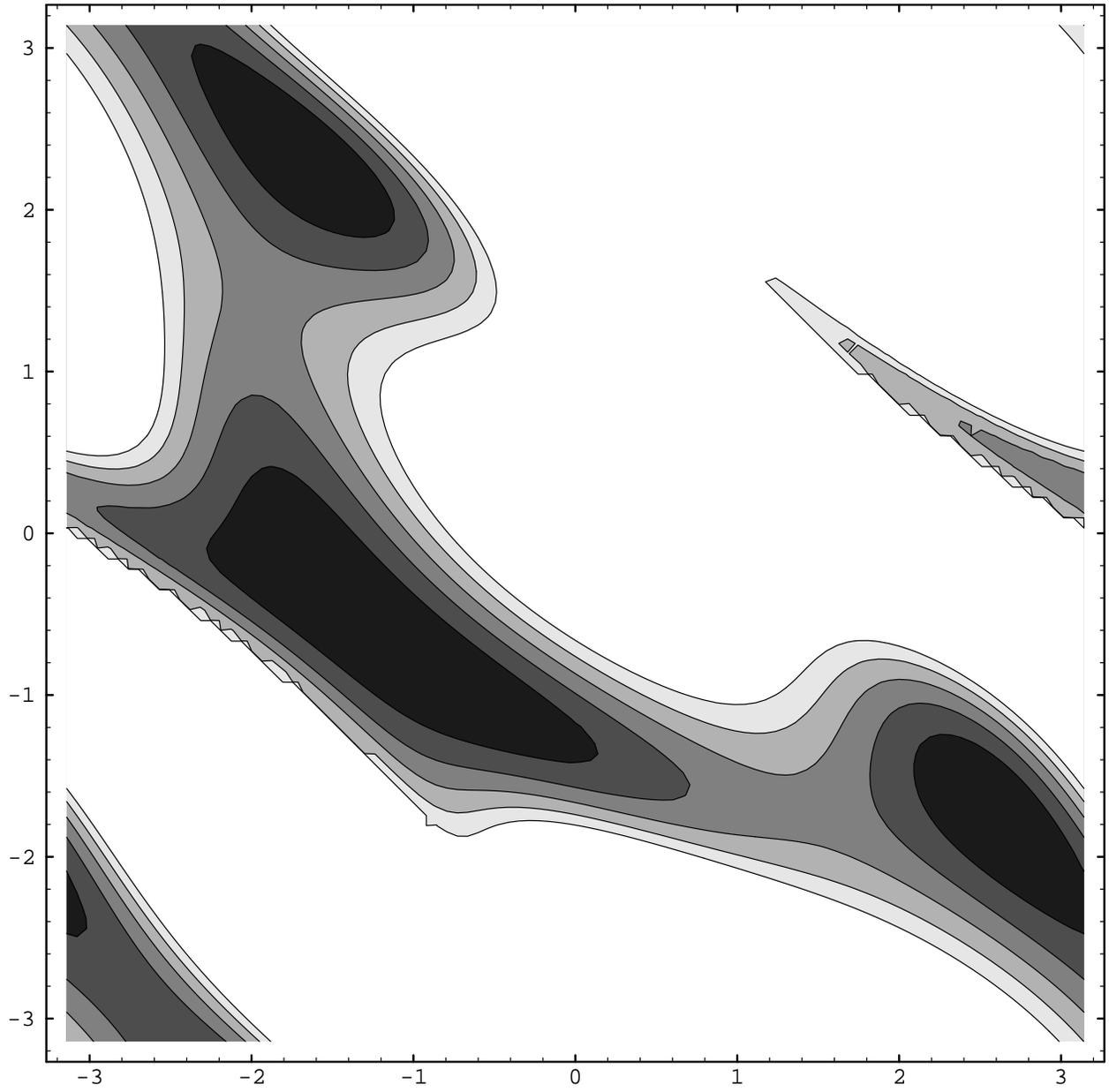
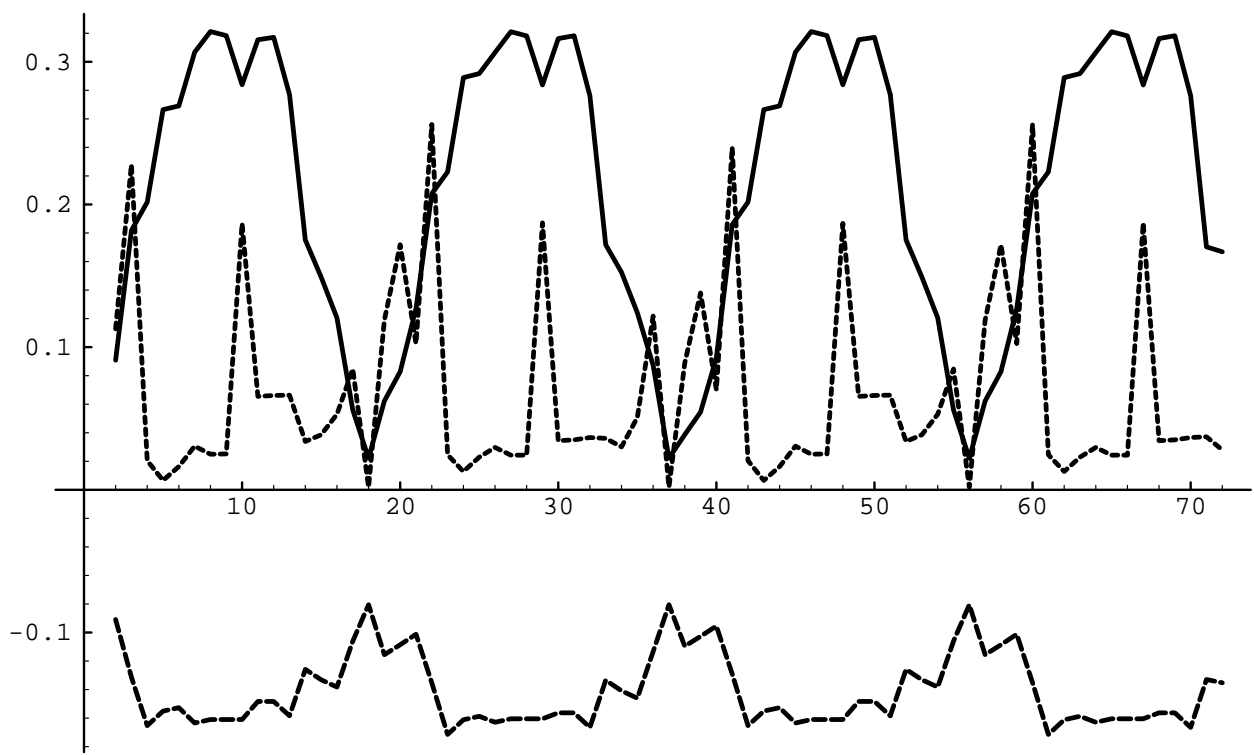
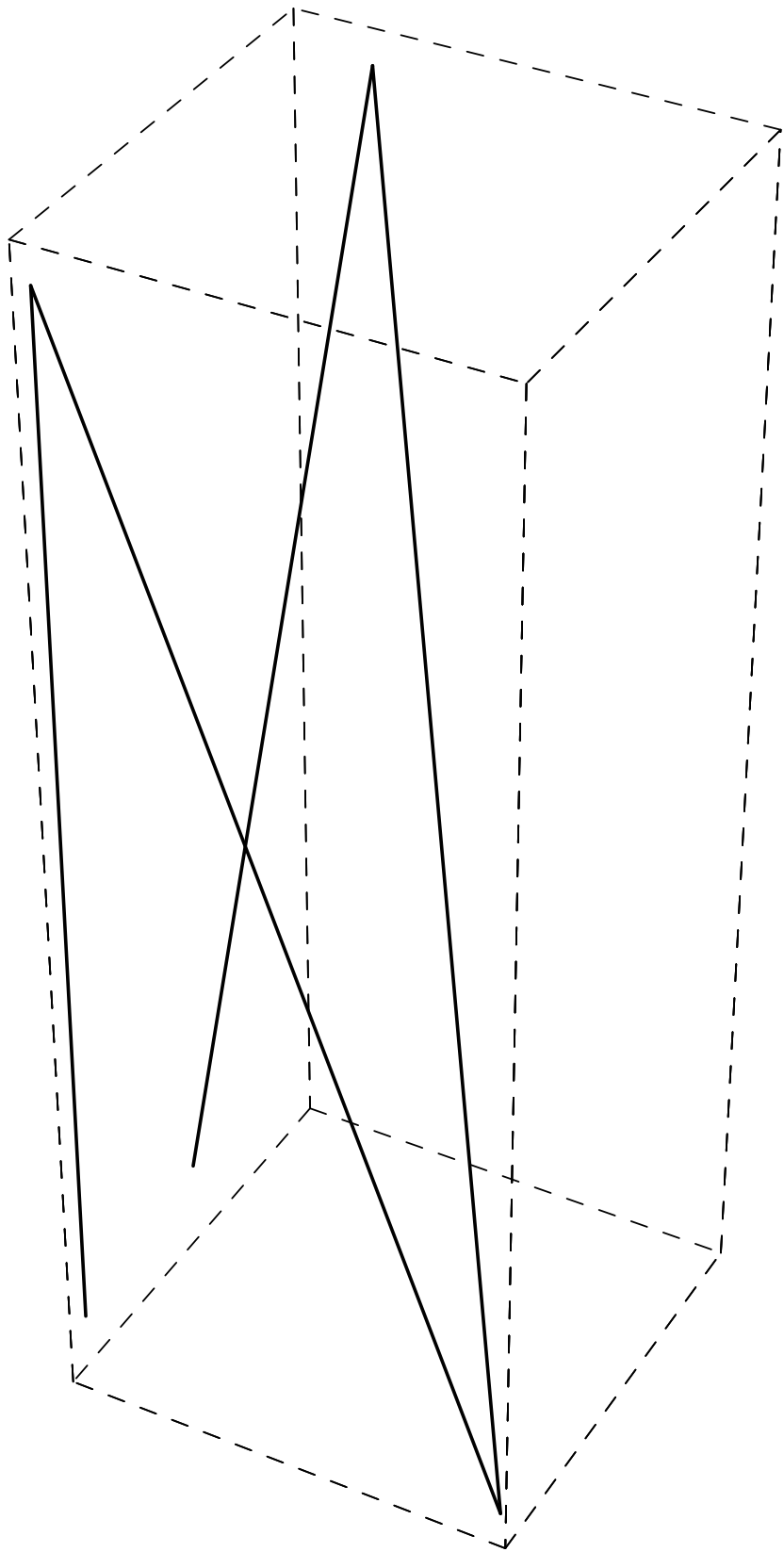
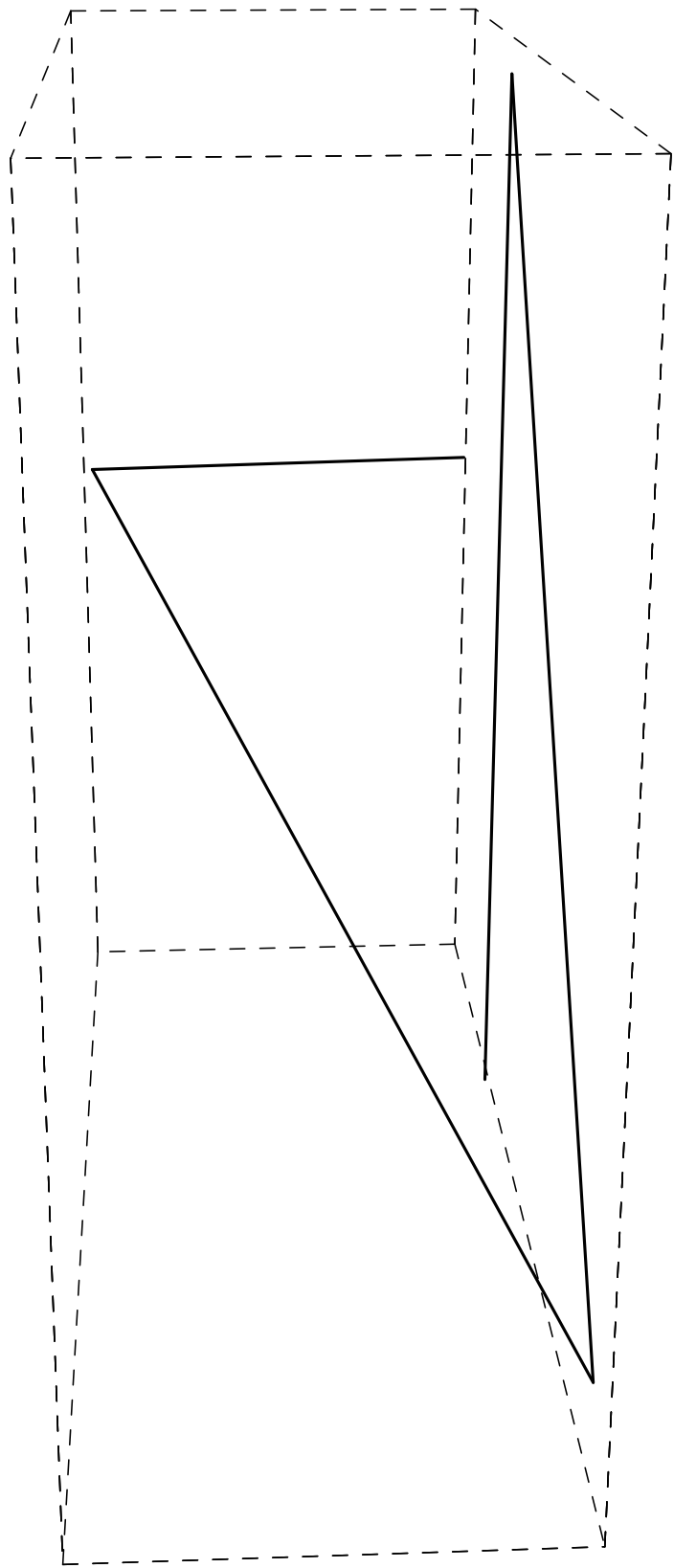
\end{document}